\documentclass[]{aastex631}

\usepackage{amsmath}

\defcitealias{Knowles2023}{Kn23}
\defcitealias{Thomas2010}{Th10}
\defcitealias{Conroy2014}{Co14}

\begin{document}

\title{\texttt{StAGE}: Stellar Archaeology-driven Galaxy Evolution}

\author[0009-0000-8215-6698]{Michele Bosi}
\affiliation{Scuola Internazionale Superiore di Studi Avanzati, Via Bonomea 265, 34136 Trieste, Italy}
\affiliation{Department of Physics, University of Trento, Via Sommarive 14, 38123 Povo (TN), Italy}

\author[0000-0002-4882-1735]{Andrea Lapi}
\affiliation{Scuola Internazionale Superiore di Studi Avanzati, Via Bonomea 265, 34136 Trieste, Italy}
\affiliation{Institute for Fundamental Physics of the Universe (IFPU), Via Beirut 2, 34014 Trieste}
\affiliation{Istituto Nazionale Fisica Nucleare (INFN), Sezione di Trieste, Via Valerio 2, 34127 Trieste, Italy}
\affiliation{INAF - Istituto di Radioastronomia, Via Gobetti 101, 40129 Bologna, Italy}

\author[0000-0003-3127-922X]{Lumen Boco}
\affiliation{Universit{\"a}t Heidelberg, Zentrum f{\"u}r Astronomie, Institut f{\"u}r theoretische Astrophysik, Albert-Ueberle-Str. 3, 69120 Heidelberg, Germany}
\affiliation{Scuola Internazionale Superiore di Studi Avanzati, Via Bonomea 265, 34136 Trieste, Italy}
\affiliation{Institute for Fundamental Physics of the Universe (IFPU), Via Beirut 2, 34014 Trieste}

\author[0009-0007-0594-1992]{Carlos Alonso-Alvarez}
\affiliation{Scuola Internazionale Superiore di Studi Avanzati, Via Bonomea 265, 34136 Trieste, Italy}

\author[0000-0003-4537-0075]{Marcos Muniz-Cueli}
\affiliation{Scuola Internazionale Superiore di Studi Avanzati, Via Bonomea 265, 34136 Trieste, Italy}
\affiliation{Institute for Fundamental Physics of the Universe (IFPU), Via Beirut 2, 34014 Trieste}

\author[0009-0005-3537-682X]{Giovanni Antinozzi}
\affiliation{Scuola Internazionale Superiore di Studi Avanzati, Via Bonomea 265, 34136 Trieste, Italy}

\author[0000-0002-6444-8547]{Meriem Behiri}
\affiliation{Scuola Internazionale Superiore di Studi Avanzati, Via Bonomea 265, 34136 Trieste, Italy}
\affiliation{INAF - Istituto di Radioastronomia, Via Gobetti 101, 40129 Bologna, Italy}

\author[0000-0002-1847-4496]{Marika Giulietti}
\affiliation{INAF - Istituto di Radioastronomia, Via Gobetti 101, 40129 Bologna, Italy}

\author[0000-0003-1186-8430]{Marcella Massardi}
\affiliation{Scuola Internazionale Superiore di Studi Avanzati, Via Bonomea 265, 34136 Trieste, Italy}
\affiliation{INAF - Istituto di Radioastronomia, Via Gobetti 101, 40129 Bologna, Italy}

\author[0000-0003-0930-6930]{Mario Spera}
\affiliation{Scuola Internazionale Superiore di Studi Avanzati, Via Bonomea 265, 34136 Trieste, Italy}
\affiliation{Istituto Nazionale Fisica Nucleare (INFN), Sezione di Trieste, Via Valerio 2, 34127 Trieste, Italy}
\affiliation{INAF - Osservatorio Astronomico di Roma, Via Frascati 33, I–00040, Monteporzio Catone, Italy}

\author[0000-0002-7922-8440]{Alessandro Bressan}
\affiliation{Scuola Internazionale Superiore di Studi Avanzati, Via Bonomea 265, 34136 Trieste, Italy}
\affiliation{Institute for Fundamental Physics of the Universe (IFPU), Via Beirut 2, 34014 Trieste}

\author[0000-0002-8211-1630]{Carlo Baccigalupi}
\affiliation{Scuola Internazionale Superiore di Studi Avanzati, Via Bonomea 265, 34136 Trieste, Italy}
\affiliation{Institute for Fundamental Physics of the Universe (IFPU), Via Beirut 2, 34014 Trieste}
\affiliation{Istituto Nazionale Fisica Nucleare (INFN), Sezione di Trieste, Via Valerio 2, 34127 Trieste, Italy}

\author[0000-0003-1186-8430]{Luigi Danese}
\affiliation{Scuola Internazionale Superiore di Studi Avanzati, Via Bonomea 265, 34136 Trieste, Italy}
\affiliation{Institute for Fundamental Physics of the Universe (IFPU), Via Beirut 2, 34014 Trieste}

\begin{abstract}
We build a semi-empirical framework of galaxy evolution (dubbed \texttt{StAGE}) firmly grounded on stellar archaeology. The latter provides data-driven prescriptions that, on a population statistical ground, allow to define the age and the star formation history for the progenitors of quiescent galaxies (QGs). We exploit \texttt{StAGE} to compute the cosmic star formation rate (SFR) density contributed by the progenitors of local QGs, and show it to remarkably agree with that estimated for high-$z$ dusty star-forming galaxies which are faint/dark in the NIR, so pointing toward a direct progenitor-descendant connection among these galaxy populations. Furthermore, we argue that by appropriately correcting the observed stellar mass density by the contribution of such NIR-dark progenitors, \texttt{StAGE} recovers a SFR density which is consistent with direct determinations from UV/IR/radio surveys, so substantially alleviating a longstanding tension. Relatedly, we also show how \texttt{StAGE} can provide the average mass and metal assembly history of QGs, and their redshift-dependent statistics. Focusing on the supermassive black holes (BHs) hosted by massive QGs, we exploit \texttt{StAGE} to reconstruct the average BH mass assembly history, the cosmic BH accretion rate density as a function of redshift, and the evolution of the Magorrian-like relationship between the relic stellar and BH masses. All in all, \texttt{StAGE} may constitute a valuable tool to understand via a data-driven, easily expandable, and computationally low-cost approach the co-evolution of QGs and of their hosted supermassive BHs across cosmic times.
\end{abstract}


\section{Introduction} \label{sec|intro}

The formation and evolution of galaxies and of their hosted supermassive black holes (BH) constitute fundamental issues in modern astrophysics and cosmology. In the local Universe, galaxies with stellar masses $M_\star\sim$ some $10^{10}-10^{12}\, M_\odot$ are mostly quiescent (QG), and feature spheroidal morphologies, old stellar populations (ages $\gtrsim$ a few Gyr), (super)solar metallicities and $\alpha$-enhanced abundances (e.g., \citealt{Thomas2005,Thomas2010,Gallazzi2006,Gallazzi2014,Wuyts2011,Johansson2012,Conroy2014,vanderWel2014,Moffett2016,Dimauro2019,Driver2022}). In the innermost regions they host a relic supermassive BH, featuring a huge mass $M_\bullet\sim 10^6-10^{10}\, M_\odot$ that intriguingly correlates with the properties of the old stellar component (e.g., \citealt{Magorrian1998,Ferrarese2000,Gebhardt2000,Tremaine2002,Kormendy2013,McConnell2013,Reines2015,Shankar2016,Sahu2019,Zhu2021}). At variance with respect to their host dark matter halos of masses $M_{\rm H}\sim 10^{11}-10^{13}\, M_\odot$ which evolve hierarchically in time by progressive merging of smaller units and/or mass accretion along filamentary structures of the cosmic web (e.g., \citealt{Springel2006,Shandarin2012,Vogelsberger2014,Vogelsberger2020,Libeskind2018,MArtizzi2019,Wilding2021,Angulo2022}), the stellar and BH mass of QGs follow a downsizing evolutionary trend: more massive galaxies (and BHs) appear to form earlier and quicker than lower mass counterparts (e.g., \citealt{Cowie1996,Thomas2005,Thomas2010,MartinNavarro2018,Merlin2019,Nanayakkara2022,Lah2023}).
A popular theoretical picture grounded on these evidences suggests local QGs to have acquired most of their stellar mass in a violent and quick star-formation episode at early epochs (redshift $z\gtrsim 1$), that after a time $\lesssim $ Gyr has been terminated by the energy/momentum feedback from the growing supermassive BH at their centers (e.g., \citealt{Granato2004,Lapi2006,Lapi2014,Lapi2018}).

This very classic framework has been reinforced in the last decade with the advent of new observational facilities in the IR, (sub-)mm, and radio bands, that have pinpointed a population of strongly star-forming galaxies (SFR $\psi\gtrsim 10^2\, M_\odot$ yr$^{-1}$), with compact morphologies (sizes $\lesssim$ a few kpc) and heavy dust obscuration (attenuation $A_V\gtrsim$ several mag) at and beyond the cosmic noon ($z\sim 2$), which have been naturally identified as the progenitors of local QGs (e.g., \citealt{Simpson2015,Straatman2016,Spilker2016,Tadaki2017,Tadaki2023,Ikarashi2015,Ikarashi2022,Fujimoto2017,Fujimoto2022,Fudamoto2022,Talia2018,Gullberg2019,Rodighiero2023,Barro2014,Barro2024}). In many instances the star formation in these progenitors is so intense that a huge amount of dust is accumulated in short timescales so as to completely obscure the emission from younger stellar populations, making these sources faint or invisible in the optical/near-IR and bright in the far-IR/(sub)-mm where dust grains reradiate the absorbed energy. Hence they have been dubbed dusty star-forming galaxies, a generic name often complemented with more specific attributes referring to their selection techniques and obscuration amount in various bands, like HST-dark, RS-NIR dark, H-dropout, NIR-dark/faint, OIR-dark, etc. (e.g., \citealt{Simpson2014,Wang2019,Sun2021,Smail2021,Talia2021,Fudamoto2021,Shu2022,Barrufet2023,Labbe2023,Perez2023,Nelson2023,vandervlugt2023,Williams2024,Gentile2024}). Multi-wavelength observations including X-ray, radio and IR/mm data have even allowed to detect the emission from the BHs growing by gas accretion at the center of these star-forming systems, and to relate its bolometric luminosity to the SFR of the host, so reinforcing the evidence of a coevolution pattern between the BH and stellar masses (e.g., \citealt{Mullaney2012,Page2012,Delvecchio2014,Delvecchio2018,Stanley2015,Stanley2017,Rodighiero2015,Rodighiero2019,Massardi2018,Combes2019,Damato2020, Laloux2023,Laloux2024,Andonie2022,Andonie2024}).

However, many details in the evolution pattern of local massive QGs are still uncertain and subjects of an intense debate in the scientific community, that curiously involves both the birth and the death of these systems. As to the former, it is somewhat unclear what has triggered the strong burst of star-formation originating most of their stellar content (and the associated accretion onto the central BH): wet galaxy mergers (e.g., \citealt{Mihos1996,Hopkins2006}), in-situ processes related to a biased collapse with large cold/molecular gas fraction or enhanced star-formation efficiency (e.g., \citealt{Bouche2010,Dave2012,Romanowsky2012,Lilly2013,Feldmann2015,Pantoni2019}), cold gas streams from large-scale
filaments of the cosmic web (e.g., \citealt{Birnboim2003,Dekel2009,Dekel2023})
violent gas instabilities driven by the fast collapse of the host dark matter halo (e.g., \citealt{Cook2009,Boco2023,Mo2004,Mo2024}), etc. As to galaxy death, the quenching of the star formation can be traced back to a variety of mechanisms, like the feedback action from stellar winds/supernova explosions and from the central BH (e.g., \citealt{Tinsley1980,White1991,Murray2005,Silk1998,Fabian1999,Granato2004,King2015rev,Lapi2006,Lapi2014,Lapi2018,Grand2017,Valentini2020,Goubert2024}), morphological quenching associated to the formation of the bulge itself (e.g., \citealt{Cook2009,Martig2009,Gensior2020}),
mass quenching via a threshold in halo mass
above which galaxies can retain a hot gas atmosphere that hinder further cold gas inflows (e.g., \citealt{Keres2005,Dekel2006}), environmental effects such as ram-pressure stripping, starvation/strangulation and gravitational heating (e.g., \citealt{Larson1980,Gunn1972,Cowie1977,Moore1999,Bekki2002,Khocfar2008,Peng2012,Roediger2014,Steinhauser2016,Man2018,Mao2022}).

On a population-wide viewpoint, a crucial issue concerns the abundance and the overall contribution of dusty star-forming galaxies to the global star formation and stellar mass assembly history of the Universe. Since QGs contain most of the stellar mass density in the local Universe, it is reasonable to expect that their star-forming progenitors could have appreciably contributed to the global SFR density. Observational estimates are still very uncertain, but recent data indicate that a sizeable fraction of the cosmic SFR density in the redshift range $z\sim 2-4$ and possibly even at higher $z$ could have occurred within dust-obscured galaxies
(e.g., \citealt{Talia2021,Behiri2023,vandervlugt2023,Williams2024,Gentile2024stat}). The interest on the subject has been further rekindled in recent years since it also lays at the crossroads of many diverse research questions in astronomy. For example, knowing the amount of stars born in dust-enshrouded environments across cosmic times and in particular at high redshift could constrain dust production models (e.g., \citealt{Dave2019,Nanni2020, Donevski2020,Parente2022,Ferrara2023,Fudamoto2024}). Moreover, dusty star-forming galaxies may also constitute sites for the birth of many stellar mass BHs, which if in binary systems may merge after some delay time to originate gravitational wave emissions. Therefore assessing the efficiency in the production of such remnants, which crucially depends on the metallicity of the host, may be relevant in forecasts for next-generation gravitational wave experiments (e.g., \citealt{Belczynski2016,Artale2019,Santoliquido2020,Boco2019,Boco2021,Broekgaarden2022,Chruslinska2019,Chruslinska2021,Chruslinska2024}). Yet another example concerns cosmological studies, since the aging with redshift of (relatively local) massive QGs may constitute an important tool to obtain a direct estimate of the Hubble parameter, independent of assumptions on the background evolution of the Universe (the so called `cosmic chronometers' method, see \citealt{Jimenez2002}; also recent review by \citealt{Moresco2024} and references therein).

Not surprisingly, the formation and evolution pattern of massive galaxies have been investigated by many theoretical models. 
There are nowadays three main approaches, that should be considered a complementary triumvirate (see \citealt{Lapi2025}): numerical simulations, semi-analytic models and semi-empirical models. Hydrodynamical simulations allow to address the galaxy formation process in fine detail, as they can cope with the simultaneous evolution of dark matter, gas and stars (e.g., \citealt{DeLucia2006,Naab2017,Dave2019,Valentini2020,Donnari2021,Habouzit2022,Piotrowska2022,Wellons2023,Feldmann2023}). However, the required long computational times limit somewhat the full exploration of the parameter space, while the thresholds in mass/volume resolution call for sub-grid recipes outside of the simulation setup to describe small-scale physical processes (e.g., dust formation, stellar and BH feedback, etc.). Semi-analytic models take up dark matter merger history from gravity-only $N-$body simulations, and describe the physics inside halos via parametric expressions gauged on (mainly) local observables (e.g., \citealt{Baugh2005,Croton2006,Monaco2007,Somerville2008,Somerville2015rev,Benson2012,Henriques2015,Henriques2020,Lacey2016,Lagos2018,Fontanot2020,Barrera2023}). They are less computationally expensive than simulations and enable to explore more rapidly the parameter space, though the high number of fudge parameters necessary to describe the complex physics may lead to degeneracies and loss of  predictive power, especially toward high redshift. In relation to the aforementioned open problems about birth, quenching, and abundance of massive QGs, all these ab-initio approaches offer some insights into the influence of the diverse physical processes, yet they struggle somewhat in reproducing the high SFRs, heavy dust content, quick mass assembly and high-redshift abundance of QG progenitors. Moreover, their outcomes are contingent on the chosen physical parameterization, sub-grid recipes and resolution (see \citealt{Scannapieco2012,Donnari2021,Crain2023}). 

Semi-empirical models try to bypass these shortcomings by adopting an ‘effective’ approach to galaxy evolution: they do not attempt to model the small-scale physics regulating the baryon cycle from first principles, but marginalize over it by exploiting empirical relations between the spatially-averaged properties of galaxies (e.g., stellar mass, SFR, specific SFR) and dark matter halos (e.g., halo mass, accretion rate, or circular velocity), derived from their relative abundance (e.g., via abundance matching) or from analytic parameterizations (e.g., \citealt{Hopkins2006,Moster2013,Moster2018,Behroozi2013,Behroozi2019,Behroozi2020,Shankar2014,Mancuso2016a,Mancuso2016b,Mancuso2017,Buchan2016,Grylls2019,Hearin2022,Drakos2022,Fu2022,Fu2024,Boco2023,Zhang2023,Zhang2024}). The value of these models stands in that they feature by design a minimal set of assumptions and parameters gauged on observations. Moreover, by empirically linking multiple observables, they can test for possible inconsistencies among distinct datasets, which can often occur given the significant observational systematics.
Finally, these frameworks are particularly helpful when galaxy formation recipes must be coupled with those from other branches of astrophysics (e.g. stellar evolution, planetary science, radiative transfer codes, etc.), and therefore it is worth to
minimize the uncertainties/hypotheses at least on the former side by exploiting basic data-driven inputs (e.g., \citealt{Lapi2024,Roy2025}). 
However, semi-empirical models are not free of downsides. By construction they strongly rely on data and, therefore, it is essential to input them with robust determinations of galaxy statistics and scaling relations between baryonic and halo properties to build a successful framework. This is somewhat challenging for sub-populations like QGs, since the related statistics (e.g., the QG stellar mass function) is uncertain especially toward high $z$ (e.g., \citealt{Moffett2016,Davidzon2017,Weaver2023}). Moreover, the association of a specific galaxy population to the subsample of dark halos hosting it is definitely nontrivial in data-driven terms (e.g., one cannot apply abundance matching). This is why semi-empirical models have been so far focusing mainly on the overall population of galaxies, and no educated treatment tailored on QGs and their star-forming progenitors exists so far. 

Here we propose to attack and solve the issue for QGs by exploiting the outcomes of stellar archaeology. The latter is a long-established technique that investigates in detail the fossil records imprinted in the stellar populations of local QGs by measuring absorption line indices and/or performing full-spectrum fitting via stellar evolution libraries (e.g., \citealt{Thomas2003,Thomas2005,Thomas2010,Gallazzi2006,Johansson2012,Gallazzi2014,Conroy2014,Worthley2014,MartinNavarro2018,Morishita2019,Saracco2020,Knowles2023,Beverage2021,Beverage2023,Beverage2024}). The outcomes of such analysis consist in prescriptions that, at least on a population statistical basis, allow to define the age and the star formation history for the progenitors of local QGs, thus obtaining clues about their stellar mass assembly across cosmic times. Despite being a classical methodology, so far it has not been systematically and quantitatively employed in models of galaxy and BH evolution. In the present work, we build the first-ever 
stellar archaeology-driven galaxy evolution framework, dubbed \texttt{StAGE}, that is aimed at reconstructing the properties, abundance and star formation history of QG progenitors and of their hosted supermassive BHs. Differently from most semi-empirical frameworks, \texttt{StAGE} does not require to link observable galaxy properties to that of the host dark matter halos, neither statistically nor parametrically, being completely agnostic about them. Moreover, \texttt{StAGE} does not require to know in advance the statistics of QGs at different redshifts, since this is self-consistently reconstructed from the local one thanks to the stellar archaeological prescriptions, so circumventing the aforementioned problems.

The plan of the paper is as follows: in Section 2 we describe our semi-empirical framework; in Section 3 we present our results; in Section 4 we discuss and summarize our findings. Throughout the paper we adopt the standard $\Lambda$CDM cosmology (\citealt{Planck2020}) and the \cite{Chabrier2003} initial mass function (IMF).

\section{Basic formalism} \label{sec|methods}

We build \texttt{StAGE}, a semi-empirical model firmly grounded on stellar archaeology. The latter has been exploited in several studies to robustly characterize the typical formation time distribution and star-formation history of (local) QGs  (e.g., \citealt{Thomas2005,Thomas2010,Gallazzi2006,Gallazzi2014,Johansson2012,Worthley2014,Conroy2014,MartinNavarro2018,Morishita2019,Saracco2020,Knowles2023,Beverage2021,Beverage2023,Beverage2024}). 
Specifically, the star formation history for the progenitors of local QGs is usually described in stellar archaeological studies with a flexible Gaussian shape\footnote{This is meant to render the typical spatially-averaged star formation histories of QG progenitors. Despite its simplicity, for different values of the width $\sigma_{\Delta t}$ it can describe rather different star-formation histories, from nearly constant (apt to small spheroids) to strongly peaked ones (apt for massive ellipticals). However, this shape could not be readily applied to other galaxy populations like disk-dominated galaxies or dwarfs, where more articulated parameterizations such as (delayed-)exponential shapes may be more appropriate; however, these are not of our concern in the present work.}:
\begin{equation}\label{eq|SFRprog}
\psi(z|M_{\star,\rm QG};t_{\rm form}) = \frac{M_{\star,\rm QG}}{1-\mathcal{R}}\, \frac{e^{-(t_z-t_{\rm form})^2/2\sigma_{\Delta t}^2}}{\sqrt{2\pi\sigma_{\Delta t}^2}}~;
\end{equation}
here $M_{\star,\rm QG}$ is the `relic' (i.e., after quenching) stellar mass, $\mathcal{R}$ is the returned gas fraction from stellar evolution (for a Chabrier IMF in the instantaneous recycling approximation $\mathcal{R}\approx 0.45$ applies), $t_z$ is the cosmic time corresponding to redshift $z$, $t_{\rm form}$ is the formation time (defined as the epoch where half of the stellar mass has been accumulated), and $\sigma_{\Delta t}$ is the star-formation duration. 

In the standard assumptions of stellar archaeology, the star-formation duration $\sigma_{\Delta t}$ is inferred from the [$\alpha/$Fe] abundance ($\alpha$-enhancement). The rationale is that $\alpha$ elements are mainly contributed by type II SN explosions and as such they are produced promptly along with the star formation, whereas iron group elements are mainly released by type I$a$ SN that explode with some delay. If the star formation is quenched before the majority of type I$a$ SN have enriched the interstellar medium with iron, an enhanced [$\alpha/$Fe] abundance locked in stars would be originated (though it would be more appropriate to speak of an iron depletion rather than of an $\alpha$-enhancement). The relation between the star formation timescale and the [$\alpha/$Fe] abundance is usually expressed as $[\alpha/{\rm Fe}]\approx \mathcal{C}-\kappa\, \log \sigma_{\Delta t}\, [{\rm Gyr}]$, where the coefficients $\mathcal{C}\approx 0.1-0.3$ and $\kappa\approx 0.1-0.2$ slightly depend on the IMF, on the star formation and chemical enrichment history, on the stellar yields, on the type I$a$ delay time distribution and on the galactic environment (e.g., \citealt{Matteucci1986,Pipino2004,Thomas2005,Romano2005,Arrigoni2010,deLaRosa2011,DeLucia2014,Vazdekis2010,Vazdekis2015,Vincenzo2016}). 

Moreover, in stellar archaeological studies the formation time distribution is routinely modeled with a log-normal shape (e.g., \citealt{Thomas2005,Thomas2010,Johansson2012,Conroy2014,Knowles2023}):
\begin{equation}\label{eq|probtform}
\frac{{\rm d}p}{{\rm d}\log t_{\rm form}}(t_{\rm form}|M_{\star,\rm QG}) = \frac{1}{\sqrt{2\pi\sigma_{\rm form}^2}}\, e^{-(\log t_{\rm form}-\log \langle t_{\rm form}\rangle)^2/2\sigma_{\log t_{\rm form}}^2}~,
\end{equation}
where $\langle t_{\rm form}\rangle=t_{0}-\langle t_{\rm age}\rangle$ is the mean formation time in terms of the present age of the Universe $t_0$ and of the mean galaxy age $\langle t_{\rm age}\rangle$, and $\sigma_{\log t_{\rm form}}\approx 0.25$ dex is the dispersion of the above distribution (corresponding to the measured dispersion in age, see \citealt{Thomas2010}). 

Stellar archaeological studies (see aforementioned references) provide the average age hence formation time $\langle t_{\rm form}\rangle(M_{\star, \rm QG})$ and the average [$\alpha/$Fe] ratio hence the star-formation duration $\sigma_{\Delta t}(M_{\star,\rm QG})$ in terms of the relic stellar mass $M_{\star,\rm QG}$ or of its proxy provided by the stellar velocity dispersion\footnote{We relate the stellar velocity dispersion to the stellar mass via the expression $\sigma_{\star,\rm QG}\, [{\rm km\, s}^{-1}]\approx 142\, \times (M_{\star,\rm QG}/10^{10.675}\, M_\odot)\, [\frac{1}{2}+\frac{1}{2}\,(M_{\star,\rm QG}/10^{10.675}\, M_\odot)^8]^{-0.227}$ provided by \cite{Cappellari2013}.} $\sigma_{\star,\rm QG}$. These relationships qualitatively agree among literature studies, although some quantitative differences are present. To explore the impact of the latter on our results, throughout the paper we will present outcomes for three different stellar archaeology prescriptions: the classic analysis by \cite[hereafter \citetalias{Thomas2010}]{Thomas2010} based on Lick indices measurements and models for a sample of individual early-type galaxies from the SDSS; the study by \cite[hereafter \citetalias{Conroy2014}]{Conroy2014} based on full spectral fitting of stacked SDSS galaxies via SPS models built with the \texttt{MILES} stellar library; the recent work by \cite[hereafter \citetalias{Knowles2023}]{Knowles2023} where Lick indices of stacked SDSS data are fitted via a SPS model built with the \texttt{sMILES} semi-empirical stellar library (including variable $[\alpha/{\rm Fe}]$ ratios). 

\begin{figure}
\epsscale{.9}\plotone{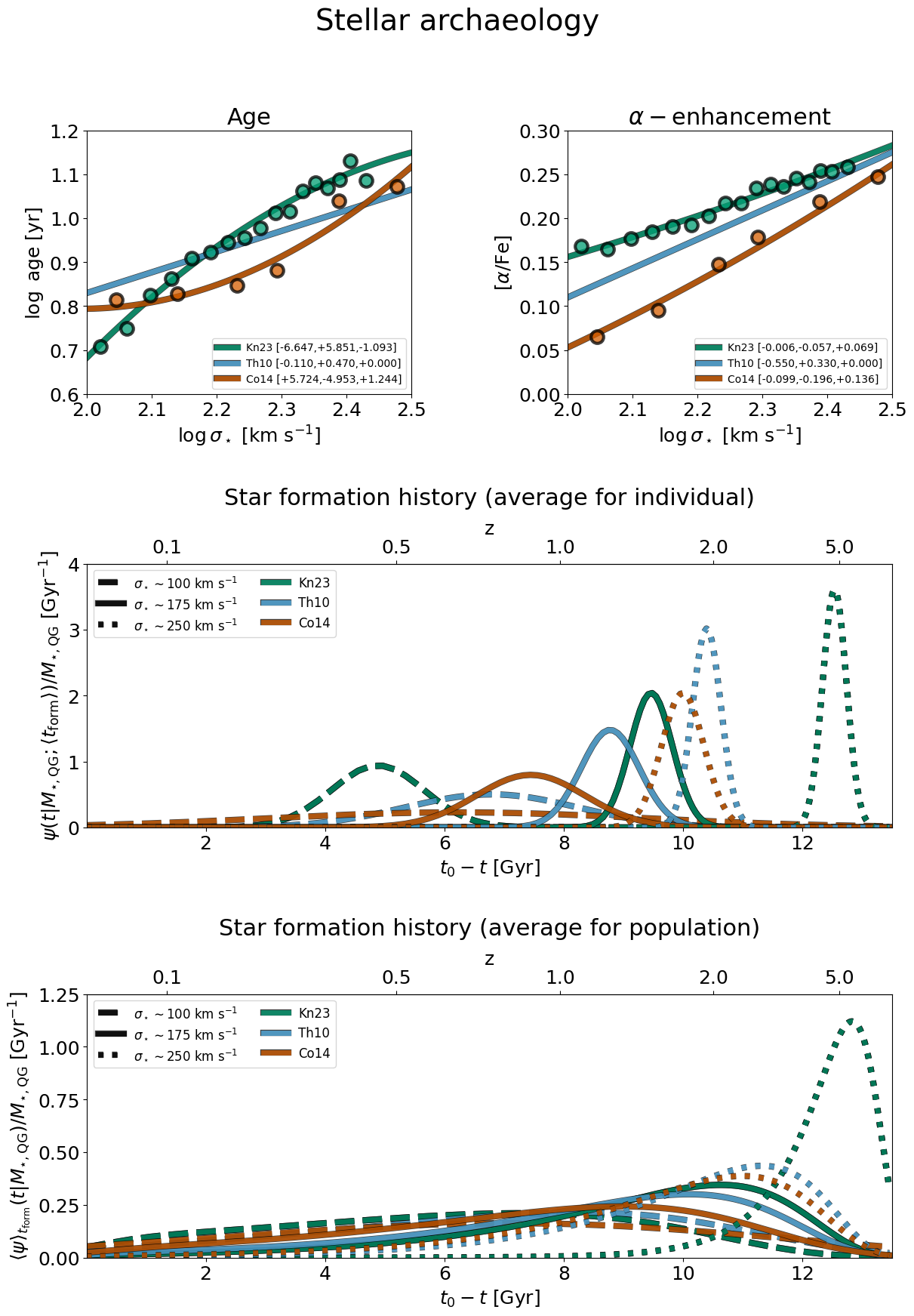}
\caption{Illustration of the stellar archaeology prescriptions employed in this work (top panel) and of the resulting star formation histories (middle and bottom panels).
Different colors refer to the stellar archaeology prescriptions by \citetalias{Knowles2023} (green), \citetalias{Thomas2010} (blue), \citetalias{Conroy2014} (red). In the top panels the age (left) or [$\alpha$/Fe] (right) vs. velocity dispersion relationships from stellar archaeology are displayed. The solid lines illustrate quadratic bestfits $X = c_0+c_1\, \log\sigma_\star+c_2\, (\log\sigma_\star)^2$, where  the coefficients $[c_0,c_1,c_2]$ are written in the legend; the original datapoints by \citetalias{Conroy2014} and \citetalias{Knowles2023} from the analysis of stacked SDSS spectra are also reported as colored circles. In the middle and bottom panels the SFR $\psi$ normalized to the relic stellar mass $M_{\star,\rm QG}$ is illustrated as a function of lookback time (lower horizontal axis) or redshift (upper horizontal axis); the middle panel refers to the average for individual galaxies (i.e., the SFR $\psi(t|M_{\star,\rm QG}; \langle t_{\rm form}\rangle)$ computed with the average formation time), the bottom panel to the population-averaged one (i.e., the SFR $\langle\psi\rangle(t|M_{\star,\rm QG})$ integrated over the formation time distribution). Dashed, solid and dotted lines refer to three different values of the stellar velocity dispersion $\sigma_\star\approx 100$, $175$ and $250$ km s$^{-1}$, corresponding to relic stellar masses spanning the range $M_{\star,\rm QG}\approx 10^{10-12}\, M_\odot$.}\label{fig|SAR}
\end{figure}

In the top panels of Figure \ref{fig|SAR} we report and compare the relationships between age and $[\alpha/{\rm Fe}]$ ratio as a function of velocity dispersion from \citetalias{Thomas2010}, \citetalias{Conroy2014} and \citetalias{Knowles2023}. In the middle panel of Figure \ref{fig|SAR} the ensuing star formation histories for different relic masses are illustrated; these constitute average renditions for individual galaxies, computed from Equation (\ref{eq|SFRprog}) when the mean formation time (or age) is employed, i.e. $\psi(z|M_{\star,\rm QG};\langle t_{\rm form}\rangle)$. Finally, in the bottom panel of Figure \ref{fig|SAR} we report the average of the star formation histories over the formation time distribution:
\begin{equation}\label{eq|psipop}
\langle\psi(z|M_{\star,\rm QG})\rangle = \int{\rm d}\log t_{\rm form}\; \frac{{\rm d}p}{{\rm d}\log t_{\rm form}}(t_{\rm form}|M_{\star,\rm QG})\, \psi(z|M_{\star,\rm QG};t_{\rm form})~.
\end{equation}
which will be relevant for population and statistical studies.
Looking at the top panel, it appears evident that the stellar archaeology relations are qualitatively similar in shape but show some quantitative difference in slope and normalization. For example, the prescription by \citetalias{Knowles2023} displays appreciably higher [$\alpha/$Fe] for all relic masses, so implying narrower star-formation histories (at given relic mass) than in the other two cases. Moreover, in terms of age vs. velocity dispersion the relation by \citetalias{Knowles2023} is steeper, to imply that galaxies with high relic masses are substantially older with respect to the other prescriptions. This behavior is reflected in the middle panel, where it can be appreciated how the star-formation histories from \citetalias{Knowles2023} are more peaked and more precocious for high relic stellar masses with respect to those from \citetalias{Thomas2010} and \citetalias{Conroy2014}. 
The population-averaged star formation histories illustrated in the bottom panel feature more similar, closely lognormal shapes, with a mode falling at $z\gtrsim 1.5$ though with an extended tail toward lower formation redshifts; this smoother behavior with respect to the corresponding histories of the middle panel is due to the average over the formation time distribution with its dispersion around $0.25$ dex. The most variant result is the population-averaged history for the \citetalias{Knowles2023} prescription at high relic masses (dotted green lines); in this case the average over formation times cannot appreciably smooth and/or skew the corresponding individual history (cf. dotted green line in the middle panel) since this is extremely precocious and narrowly peaked due to the very large [$\alpha/$Fe] and age implied at high masses by the \citetalias{Knowles2023} prescription. But apart from such an extreme instance, the relative similarity of the population-averaged star-formation histories already highlights that one is not to expect substantial difference in the outcomes of the various stellar archaeology prescriptions when employing them in galaxy evolution studies, as we will also show explicitly throughout this paper.

\subsection{The cosmic SFR density of QG progenitors}\label{sec|SFRD}

One fundamental quantity we aim to obtain is the amount of cosmic SFR density that has been contributed at a given redshift $z$ by the progenitors of galaxies which are quiescent at a lower $z_{\rm QG}<z$ (e.g., $z_{\rm QG} =0$ can be the present time). Relying on the above stellar archaeological prescriptions, this can be just expressed as
\begin{align}\label{eq|SFRDqui}
\nonumber\rho_{\rm SFR}(z|z_{\rm QG}) &= \int{\rm d}\log M_{\star,\rm QG}\, \frac{{\rm d}N_{\rm QG}}{{\rm d}\log M_{\star,\rm QG}\, {\rm d}V}\times\\
\\
\nonumber&\times\int_{-\infty}^{\log t_{z_{\rm QG}}}{\rm d}\log t_{\rm form}\, \frac{{\rm d}p}{{\rm d}\log t_{\rm form}}(t_{\rm form}|M_{\star,\rm QG})\, \psi(z|M_{\star,\rm QG};t_{\rm form})~.
\end{align}
Here ${\rm d}N_{\rm QG}/{\rm d}\log M_{\star,\rm QG}\, {\rm d}V$ is the galaxy stellar mass function of local quiescent galaxies with relic stellar mass $M_{\star,\rm QG}$; we adopt the recent determination by \cite{Weaver2023}. Moreover, the quantity $\psi(z|M_{\star,\rm QG};t_{\rm form})$ is the SFR at redshift $z$ of the star-forming progenitor for a QG with final stellar mass $M_{\star,\rm QG}$ at $z_{\rm QG}$ and formation time $t_{\rm form}$ expressed by Equation (\ref{eq|SFRprog}), while ${\rm d}p/{\rm d}\log t_{\rm form}$ is the formation time distribution from Equation (\ref{eq|probtform}) in turn depending on $M_{\star,\rm QG}$. 

For several applications in astronomy what one really needs is the SFR density sliced in redshift and stellar mass or metallicity bins. For deriving the former, a preliminary step is to compute the evolution with redshift of the stellar mass for a QG progenitor that will end up with a relic stellar mass $M_{\star,\rm QG}$ at $z_{\rm QG}$. This can be obtained just by integrating in time the star-formation history expressed by Equation (\ref{eq|SFRprog}), to yield
\begin{equation}\label{eq|Mstarprog}
M_{\star}(z|M_{\star,\rm QG};t_{\rm form}) = \int{\rm d}t\; \psi(z|M_{\star,\rm QG};t_{\rm form})=\frac{M_{\star,\rm QG}}{2}\,
\left[1+{\rm erf}\left(\frac{t_z-t_{\rm form}}{\sqrt{2}\sigma_{\Delta t}}\right)\right]~.
\end{equation}
Then the cosmic SFR density per stellar mass bin (from QG progenitors) reads
\begin{equation}\label{eq|SFRDMstarprog}
\frac{{\rm d}\rho_{\rm SFR}}{{\rm d}\log M_{\star}}  = \int{\rm d}\log M_{\star,\rm QG}\,\frac{{\rm d}N_{\rm QG}}{{\rm d}\log M_{\star,\rm QG}\, {\rm d}V}\, \int{\rm d}\log t_{\rm form}\, \frac{{\rm d}p}{{\rm d}\log t_{\rm form}}\, \psi(z)\, \delta_{\rm D}\Big[\log M_{\star}-\log M_{\star}(z)\Big]~,
\end{equation}
where $\delta_{\rm D}[\cdot]$ indicates the Dirac $\delta$-function, while $\psi(z)=\psi(z|M_{\rm QG};t_{\rm form})$ and $M_\star(z)=M_{\star}(z|M_{\rm QG};t_{\rm form})$ are given by Equations (\ref{eq|SFRprog}) and (\ref{eq|Mstarprog}). The rationale of this expression is to select from Equation (\ref{eq|SFRDqui}) only progenitor galaxies in a certain bin of stellar mass $M_\star$. In addition, the number density of QG progenitors in stellar mass bins ${\rm d}N/{\rm d}\log M_\star\,{\rm d}V$ can be derived from the above expression by removing $\psi(z)$ from the inner integral.

Now that both the SFR and the stellar mass of the progenitors are known, one can assign a (gas) metallicity $Z$ to them. In the spirit of a data-driven model we will base on observations, which indicate that the metallicity distributions of star-forming galaxies can be described as a log-normal shape around the fundamental metallicity relation (FMR; see \citealt{Mannucci2010,Mannucci2011,Andrews2013,Curti2020,Curti2023}):
\begin{equation}\label{eq|probzeta}
\frac{{\rm d}p}{{\rm d}\log Z}(Z|M_\star,\psi,z) = \frac{1}{\sqrt{2\,\pi\,\sigma_{\rm FMR}^2}}\, e^{-[\log Z-\log Z_{\rm FMR}]^2/2\sigma_{\rm FMR}^2} ~,
\end{equation}
where $\log Z_{\rm FMR}(M_\star,\psi, z)$ is the mean FMR and $\sigma_{\rm FMR}\approx 0.1$ dex is 
the measured dispersion around it. We exploit the FMR by \cite{Andrews2013} that has been shown to reproduce very well recent JWST data up to high redshifts (see \citealt{Nakajima2023}). Thus the cosmic SFR density per metallicity bin (from QG progenitors) reads
\begin{equation}\label{eq|SFRDZprog}
\frac{{\rm d}\rho_{\rm SFR}}{{\rm d}\log Z}  = \int{\rm d}\log M_{\star,\rm QG}\, \frac{{\rm d}N_{\rm QG}}{{\rm d}\log M_{\star,\rm QG}\, {\rm d}V}\, \int{\rm d}\log t_{\rm form}\, \frac{{\rm d}p}{{\rm d}\log t_{\rm form}}\, \frac{{\rm d}p}{{\rm d}\log Z}\, \psi(z)~,
\end{equation}
where ${\rm d}p/{\rm d}\log Z$ is given by Equation (\ref{eq|probzeta}) with mean value $Z_{\rm FMR}[M_\star(z),\psi(z),z]$ computed in terms of the progenitor stellar mass $M_\star(z)=M_{\star}(z|M_{\rm QG};t_{\rm form})$ and SFR $\psi(z)=\psi(z|M_{\rm QG};t_{\rm form})$. The rationale of this expression is to select from Equation (\ref{eq|SFRDqui}) only progenitor galaxies in a certain bin of metallicity according to the probability distribution of Equation (\ref{eq|probzeta}).

Finally, it can be of some interest to compute the cosmic stellar mass density of the galaxies which will be quiescent at $z_{\rm QG}$. This can be expressed as an integral over cosmic time of the SFR density, and exchanging the order of integrations one immediately finds
\begin{align}
\nonumber\rho_{\star}(z|z_{\rm QG}) & = \int{\rm d}z\, \left|\frac{{\rm d}t}{{\rm d}z}\right|\, \rho_{\rm SFR,QG}(z|z_{\rm QG}) =\\
\\
\nonumber& = \int{\rm d}\log M_{\star,\rm QG}\,\frac{{\rm d}N_{\rm QG}}{{\rm d}\log M_{\star,\rm QG}\, {\rm d}V}\, \int{\rm d}\log t_{\rm form}\, \frac{{\rm d}p}{{\rm d}\log t_{\rm form}}\, M_\star(z|M_{\star,\rm QG};t_{\rm form})~;
\end{align}
plainly at $z=z_{\rm QG}$ where $M_\star(z)=M_{\star,\rm QG}$, the integral over the formation time distribution equals unity and this expression boils down to the stellar mass density of quiescent galaxies.

\subsection{The average mass assembly of QG progenitors}\label{sec|AHQG}

\texttt{StAGE} can be exploited to study the average mass and metal assembly history of local QGs. In this respect, a crucial quantity is the 
stellar mass averaged over the formation time distribution for a galaxy that will be quiescent at present with relic mass $M_{\star,\rm QG}$. This is given by
\begin{equation}
\langle M_\star\rangle (z|M_{\star,\rm QG}) = \int{\rm d}\log t_{\rm form}\, \frac{{\rm d}p}{{\rm d}\log t_{\rm form}}(t_{\rm form}|M_{\star,\rm QG})\, M_\star(z|M_{\star,\rm QG};t_{\rm form})~;
\end{equation}
which is essentially Equation (\ref{eq|Mstarprog}) averaged over the formation time distribution from stellar archaeology expressed by Equation (\ref{eq|probtform}). If interested in chemical enrichment, one can compute the stellar metallicity, which is obtained by integrating the gas metallicity over the star formation history. In this case the computation to be performed is slightly more complex, and reads
\begin{align}\label{eq|metallicity}
\nonumber \langle Z_{\star}\rangle (z|M_{\star,\rm QG}) &= \int{\rm d}\log t_{\rm form}\, \frac{{\rm d}p}{{\rm d}\log t_{\rm form}}(t_{\rm form}|M_{\star,\rm QG})\times\\
&\\
\nonumber &\times \frac{1}{M_{\star}(z)}\, \int^z{\rm d}z'\, \left|\frac{{\rm d}t}{{\rm d}z'}\right|\, \psi(z')\, Z_{\rm FMR}\Big[M_\star(z'),\psi(z'),z'\Big]\, 10^{\sigma_{\rm FMR}^2/2}~.
\end{align}
Here $\psi(z)=\psi(z|M_{\rm QG};t_{\rm form})$ and $M_\star(z)=M_{\star}(z|M_{\rm QG};t_{\rm form})$ are given by Equations (\ref{eq|SFRprog}) and (\ref{eq|Mstarprog}). The quantity $Z_{\rm FMR}\, 10^{\sigma_{\rm FMR}^2/2}$ is the mean metallicity extracted from the (lognormal) metallicity distribution of Equation (\ref{eq|probzeta}) centered on the FMR; this is first averaged over the star-formation history (inner integral) and then over the formation time distribution (outer integral).

\subsection{Supermassive black holes evolution}\label{sec|BHARD}

\texttt{StAGE} is also suited to investigate the evolution of supermassive BHs, which are hosted at the center of massive QGs in the local Universe, and have accreted most of their mass within QG star-forming progenitors toward high redshift. In particular, we can easily compute the cosmic accretion rate density of supermassive BH in the progenitor of local QGs (which is the analog of the cosmic SFR density for BHs). Explicitly, we can write:
\begin{align}\label{eq|BHARDqui}
\nonumber \rho_{\rm BHAR}(z|z_{\rm QG}) &= \int{\rm d}\log M_{\star,\rm QG}\, \frac{{\rm d}N_{\rm QG}}{{\rm d}\log M_{\star,\rm QG}\, {\rm d}V}\, \int{\rm d}\log t_{\rm form}\, \frac{{\rm d}p}{{\rm d}\log t_{\rm form}}(t_{\rm form}|M_{\star,\rm QG})\times \\
\\
\nonumber &\times \int{\rm d}\log\lambda_X\, \frac{{\rm d}\delta}{{\rm d}\log \lambda_X}[\lambda_X|M_\star(z),z]\, \dot M_\bullet[\lambda_X,M_\star(z)] ~,
\end{align}
Here $\lambda_X\equiv L_X/M_\star$ is a pseudo-Eddington ratio given by the ratio of the X-ray luminosity $L_X$ to the progenitor stellar mass $M_\star(z)=M_\star(z|M_{\star,\rm QG},t_{\rm form})$, and ${\rm d}\delta/{\rm d}\log \lambda_X$ is the duty cycle per bin of $\lambda_X$ at given stellar mass and redshift. The above expression is useful because the latter quantity has been determined observationally from multi-wavelength AGN surveys (e.g.,  \citealt{Bongiorno2012,Georgakakis2017,Aird2018,Yang2018}); we adopt the determination by \cite{Yang2018} that have been shown to be consistent with the stellar mass function and with the AGN X-ray luminosity functions. Then the BH accretion rate follows as $\dot M_\bullet = k_X\, L_X\, (1-\epsilon)/\epsilon\, c^2$ in terms of the BH radiative efficiency $\epsilon\approx 0.1$, and of the X-ray (luminosity-dependent) bolometric corrections $k_X$ that we take from \cite{Duras2020}. Plainly avoiding to integrate over the stellar mass or over the pseudo-Eddington ratio yields the BH accretion rate density per bin of these quantities.
Finally, removing the BH accretion rate $\dot M_\bullet$ from the inner integral will provide the number density of galaxies hosting accreting supermassive BHs as a function of redshift. 

We stress one relevant difference of our framework with respect to other semi-empirical, data-driven models of galaxy-BH coevolution (e.g., \citealt{Shankar2013,Aversa2015,Carraro2022,Sicilia2022}). 
Such approaches derive the cosmic BH accretion rate using as a basic input the true Eddington ratio $\lambda= 
k_X L_X/M_\bullet$ distribution. The latter is typically derived from the observed pseudo-Eddington ratio $\lambda_X= L_X/M_\star$ by assuming a bolometric correction $k_X$ and a specific Magorrian-like $M_\bullet-M_\star$ relationship; the local $M_\bullet-M_\star$ is routinely taken as a reference, in some cases adding a parameterized redshift evolution. However, this is a strong assumption, since the shape and redshift evolution of the $M_\bullet-M_\star$ is uncertain and strongly debated. On the contrary, \texttt{StAGE} exploits stellar archaeology to reconstruct the redshift-dependent statistics of QGs at different redshift, and can incorporate directly the observed pseudo-Eddington ratio distribution as a function of redshift and stellar mass into the computation of Equation (\ref{eq|BHARDqui}). The Magorrian-like relation $M_\bullet-M_\star$ as a function of redshift is thus not assumed, but actually an output of our model.

\section{Results}\label{sec|results}

We now exploit the formalism described in Section \ref{sec|methods} to address some interesting issues in galaxy and BH evolution. 

\subsection{The SFR and stellar mass density of QG progenitors}

\begin{figure}
\epsscale{0.9}\plotone{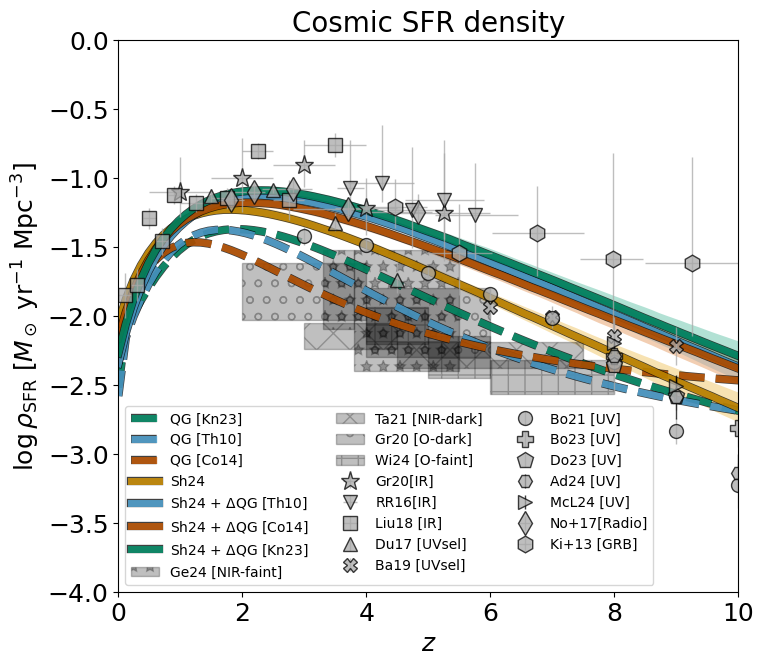}
\caption{The cosmic star formation rate density as a function of redshift. Data referring to the overall SFR density are from \citet[stars]{Gruppioni2020}, \citet[inverse triangles]{Rowan2016}, \citet[squares]{Liu2018}, \citet[triangles]{Dunlop2017}, \citet[crosses]{Bhatawdekar2019}, \citet[circles]{Oesch2018,Bouwens2021}, \citet[plus signs]{Bouwens2023}, \citet[pentagons]{Donnan2023}, \citet[rotated hexagons]{Adams2024}, \citet[rotated triangles]{McLeod2024}, \citet[rhomboids]{Novak2017}, and \citet[hexagons; from GRB]{Kistler2013}. Data referring to high-$z$ dusty galaxies are from \citet[gray shaded area hatched with circles; IR-selected HST-dark sources]{Gruppioni2020}, \citet[gray shaded area hatched with crosses; radio-selected NIR-dark sources]{Talia2021}, \citet[gray shaded area hatched with stars; radio-selected NIR-faint sources observed with JWST]{Gentile2024stat}, \citet[gray shaded area hatched with plus signs; O-dark sources observed with JWST]{Williams2024}. The orange line with shade illustrates a log-normal fit (median and $1\sigma$ uncertainty) to the SFR density that corresponds to the mass density measured by \citet[solid orange line in Figure \ref{fig|SMD}]{Shuntov2025}. The green, blue and red lines illustrate the outcomes of \texttt{StAGE} when the stellar archaeology prescriptions by \citetalias{Knowles2023}, \citetalias{Thomas2010} or \citetalias{Conroy2014} are employed, respectively: dashed lines display the SFR density associated to the progenitors of local QGs; solid lines with shades are lognormal fits to the SFR density that correspond to the mass density (solid green, blue and red lines in Figure \ref{fig|SMD}) obtained when adding the contribution of star-forming progenitors from local QGs to the measurements by \citet[see Section \ref{sec|results} for details]{Shuntov2025}.}\label{fig|SFRD}
\end{figure}

Figure \ref{fig|SFRD} illustrates the cosmic SFR density as a function of redshift. We report a comprehensive compilation of data, both those referring to the overall SFR density (symbols) measured via UV, IR/mm and radio indicators, and those referring to the contribution from dusty star-forming galaxies at $z\sim 3-5$ (hatched areas). The outcome of \texttt{StAGE} for the progenitors of QGs is highlighted by the dashed lines, with the different colors referring to the stellar archaeology prescriptions by \citetalias{Knowles2023} (green), \citetalias{Thomas2010} (blue) and \citetalias{Conroy2014} (red). \texttt{StAGE} predicts that the contribution from the star-forming progenitors of local QGs to the overall SFR density can be appreciable, amounting to several tens percent around the cosmic noon at $z\sim 2$. At $z\sim 3-5$ the prediction of \texttt{StAGE} remarkably agrees with the contribution to the SFR density estimated for dusty star-forming galaxies which are faint or invisible in the optical/NIR. This occurrence indicates that such systems can constitute the star-forming progenitors of local massive QGs, as already tentatively suggested in the literature both from a theoretical and an observational perspective (see references in Section \ref{sec|intro}). Here such a suggestion is confirmed in an independent, data-driven way, just relying on stellar archaeology.

\begin{figure}
\epsscale{0.7}\plotone{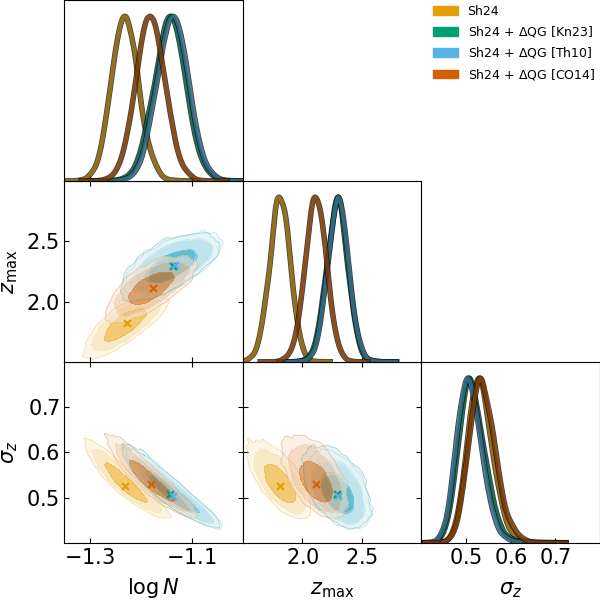}
\caption{MCMC posterior distributions for a fit to the SFR density vs. redshift with lognormal shape $\rho_{\rm SFR}(z)=N\,e^{-\ln^2[(1+z)/(1+z_{\rm max})]/2\sigma_z^2}$. Orange contours and lines refer to the SFR density that corresponds to the mass density measured by \citet[solid orange line in Figure \ref{fig|SMD}]{Shuntov2025}. The green, blue, and red lines illustrate the outcomes of \texttt{StAGE} when the stellar archaeology prescriptions by \citetalias{Knowles2023}, \citetalias{Thomas2010} or \citetalias{Conroy2014} are employed, respectively. Specifically these refer to the SFR density that corresponds to the mass density (solid red, blue and cyan lines in Figure \ref{fig|SMD}) obtained when adding the contribution of star-forming progenitors of local QGs to the measurements by \cite{Shuntov2025}. The contours show 68\%, 95\% and 99\% confidence intervals, the crosses display the maximum likelihood positions, and the marginalized distributions are in arbitrary units (normalized to 1 at their maximum value).}\label{fig|SFRD_corner}
\end{figure}

\begin{figure}
\epsscale{1}\plotone{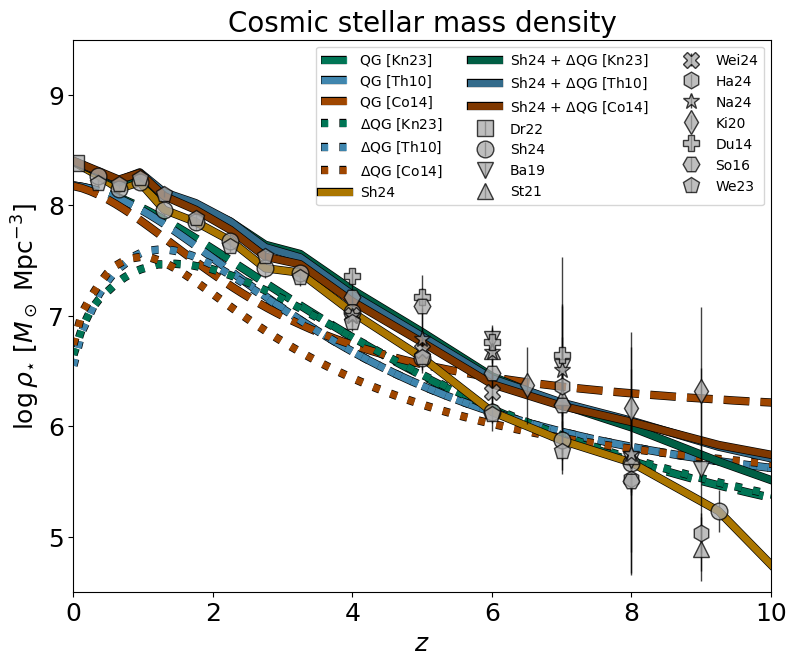}
\caption{The cosmic stellar mass density as a function of redshift (integrated over stellar masses $M_\star\gtrsim 10^8\, M_\odot$). Data are from \citet[squares]{Driver2022}, \citet[circles]{Shuntov2025}, \citet[inverse triangles]{Bhatawdekar2019}, \citet[triangles]{Stefanon2021}, \citet[crosses]{Weibel2024}, \citet[hexagons]{Harvey2025}, \citet[stars]{Navarro-Carrera2024}, \citet[rhomboids]{Kikuchihara2020}, \citet[plus signs]{Duncan2014}, \citet[rotated hexagons]{Song2016} and \citet[pentagons]{Weaver2023}. The solid orange line just connects the datapoints by \cite{Shuntov2025}.
The green, blue and red lines illustrate the outcomes of \texttt{StAGE} when the stellar archaeology prescriptions by \citetalias{Knowles2023}, \citetalias{Thomas2010}
or \citetalias{Conroy2014} are employed, respectively: the dashed lines display the stellar mass density associated to the progenitors of local QGs; the dotted lines represent the contribution from progenitors that are currently (i.e., at the given redshift $z$) star-forming and had at the peak of their activity a specific sSFR$\gtrsim 0.1$ Gyr$^{-1}$; the solid lines add the latter contributions to the orange line.}\label{fig|SMD}
\end{figure}

There is another interesting issue that the outcomes of \texttt{StAGE} can help in clearing up. It concerns the well-known discrepancy between the SFR density inferred from direct indicators, and that corresponding to the stellar mass density measured from SED fitting (mainly of O-NIR data). To explain the point, it is worth looking in parallel at the SFR and at the stellar mass density. Figure \ref{fig|SMD} illustrates the cosmic stellar mass density in galaxies with $M_\star\gtrsim 10^8\, M_\odot$ as a function of redshift, as measured from NIR surveys (datapoints in the Figure). The orange solid line passes through the most recent datapoints from the COSMOS-Web survey by \cite{Shuntov2025}. One can straightforwardly compute the SFR density that corresponds to such a stellar mass density, since the two quantities are just related by $\rho_\star(t) = (1-\mathcal{R})\, \int{\rm d}t'\, \rho_{\rm SFR}(t')$. We assume a lognormal parameterization of the overall SFR density $\rho_{\rm SFR}(z)=N\,e^{-\ln^2[(1+z)/(1+z_{\rm max})]/2\sigma_z^2}$ in terms of three parameters (the normalization $N$, the redshift $z_{\rm max}$ where the SFR attains its maximal value, and the width $\sigma_z$) and we infer these by fitting the stellar mass density estimate of \cite{Shuntov2025} via a standard MCMC algorithm, to obtain the constraints reported in Figure \ref{fig|SFRD_corner} (orange contours and lines) and in Table \ref{tab|MCMC}. The resulting SFR density is plotted in Figure \ref{fig|SFRD} as an orange line (surrounded by a shaded area that displays the fit $1\sigma$ confidence interval), which clearly shows a tension with the direct SFR measurements. In particular, it peaks at $z\lesssim 2$ which appears to be slightly lower than indicated by the data, and features a normalization that falls short with respect to the direct IR/mm estimates of the SFR density, especially in the range $z\sim 2-5$.

Although part of this discrepancy may be attributed to systematic uncertainties in the direct SFR density estimates, the outcomes of \texttt{StAGE} suggest another intriguing possibility. Given that the contribution to the SFR density from the progenitors of local QGs agrees with that of NIR-dark sources at redshifts $z\sim 2-5$, it could be that this contribution is partly missed in the stellar mass density estimates from NIR surveys. The dashed lines in Figure \ref{fig|SMD} represent the stellar mass density from the progenitors of local QGs as predicted by \texttt{StAGE}. One could be tempted just to add this contribution to the stellar mass density measured by \cite{Shuntov2025} and then refit for the corresponding SFR density. However, this would not be correct. In fact, in the mass density produced by \texttt{StAGE} at different redshifts all progenitors of local QGs are included: a galaxy progenitor at a given redshift could be still star-forming or already quiescent; moreover, depending on its formation time and star-formation duration it could have had a very bursty and peaked star-formation history or a very quiet and prolonged one. It is reasonable to think that only objects that had a very peaked and bursty star formation history have been able to accumulate huge amounts of dust so as to appear very faint or dark in the NIR. Moreover, these objects should be required to still retain a decent amount of dust, and so to be star-forming at the time of observations in order to escape the NIR detection or in order for their masses to be underestimated in the NIR-based determinations. To select these objects, we exploit \texttt{StAGE} to compute the stellar mass density contributed by the progenitors of local QGs that had at the peak of their star formation history a specific sSFR$\equiv \psi/M_\star\gtrsim $ Gyr$^{-1}$ (a criterion often adopted to define a `bursty' star-formation) and to be still star-forming at the redshift of observation with sSFR$\gtrsim 0.1$ Gyr$^{-1}$ (a threshold often adopted to discriminate between star-forming and quenched galaxies). In practice, this is done inserting these conditions as Heaviside functions in the integrand of Equation (\ref{eq|SFRDqui}). 

The outcome is illustrated in Figure \ref{fig|SMD} by the dotted lines. Given the prescriptions from stellar archaeology, such galaxies with a bursty star-formation history in the past are mainly massive local QGs, that tend to be already in passive evolution at $z\lesssim 1.5$, explaining why their contribution to the stellar mass density decreases in that redshift range. Thus we now can confidently add such contribution to the mass density from NIR data as measured by \cite{Shuntov2025}, to obtain the green, blue and red lines (depending on the stellar archaeology prescription employed). It is evident that such a mass density corrected for the contribution of NIR-dark sources is negligibly changed at $z\lesssim 2$ but it is somewhat increased toward higher redshifts. We have refitted the SFR density that corresponds to this corrected stellar mass density: the new posterior distributions and marginalized parameter estimates are illustrated in Figure \ref{fig|SFRD_corner} and reported in Table \ref{tab|MCMC} for the three different stellar archaeological prescriptions employed in this work. The resulting SFR density is illustrated in Figure \ref{fig|SFRD} (solid green, blue and red lines) and is seen to be characterized by a higher normalization and a broader peak slightly shifted toward high $z$. Overall, these outcomes agree considerably better with direct SFR density estimation via IR/mm/radio surveys, especially in the redshift range $z\sim 2-5$. At very high redshift, this corrected determination is slightly higher than current estimates based on UV data, pointing toward the putative presence of a non-negligible amount of obscured sources, as sporadically found by JWST (e.g., \citealt{Alvarez2023,Ling2024,Williams2024}); future surveys and follow-up observations in the mm/radio band with SKAO (see \citealt{Norris2015}) and with ALMA after its wide-band sensitivity upgrade (see \citealt{Carpenter2023}) will be aimed at pinpointing them and clear any residual mismatch.

\begin{figure}
\epsscale{1.}\plotone{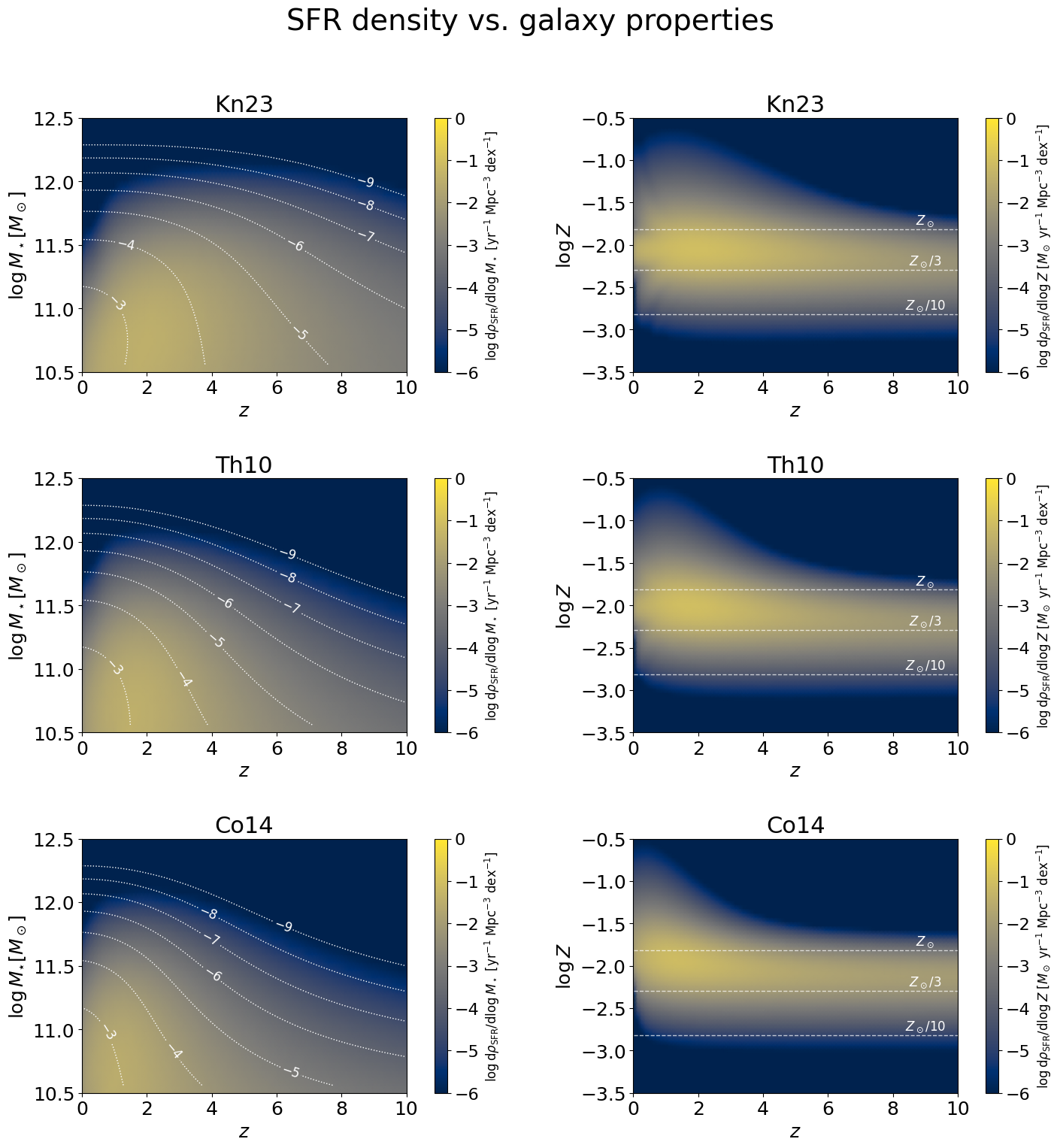}
\caption{The cosmic SFR density (color-coded) sliced in galaxy stellar mass (left panels) and galaxy metallicity (right panels) as a function of redshift, as obtained from \texttt{StAGE}. Top panels refer to the stellar archaeology prescriptions by \citetalias{Knowles2023}, middle panels to those by \citetalias{Thomas2010} and bottom panels to those by \citetalias{Conroy2014}. In the left panels the white contours with labels display the galaxy number densities per stellar mass bin $\log [{\rm d}N/{\rm d}\log M_\star\, {\rm d}V]$, in logarithmic units of Mpc$^{-3}$ dex$^{-1}$. In the right panels the three horizontal dashed lines highlight the metallicity values: $Z_\odot$, $Z_\odot/3$ and $Z_\odot/10$ from top to bottom.}
\label{fig|SFRD_2D}
\end{figure}

Figure \ref{fig|SFRD_2D} illustrates the SFR density for the progenitors of local QGs sliced in terms of their  stellar mass and metallicity, as a function of redshift. 
The plots highlight that most of the SFR density is contributed by progenitors with metallicity in the range $Z/Z_\odot\sim 1/3-1$, and even supersolar values toward lower $z$. At high redshifts, typical metallicities around a few tenths solar apply, and only for a minor fraction of the population they fall below $Z_\odot/10$ at $z\gtrsim 4$. This is due to the fact that, according to stellar archaeology prescriptions, galaxies with a very early formation time are typically more massive and characterized by a peaked and rather short star-formation history, so their chemical enrichment is quite fast and substantial. In terms of stellar masses, it is interesting to note that at the cosmic noon $z\sim 2$ a substantial contribution to the SFR density comes from already massive objects $M_\star\sim 10^{11}-10^{12}\, M_\odot$, and that an appreciable number of them is also contributing for $z\gtrsim 4$. The white contours with labels display the number density of galaxies per stellar mass bins in log units of Mpc$^{-3}$ dex$^{-1}$. The different shape of the color pattern with respect to the contours reflects that the former includes the weighting by the SFR with its mass and redshift dependence (cf. Section \ref{sec|SFRD}). Although more massive galaxies tend to be less numerous, they also feature higher SFRs, so that their contribution to the SFR density of the Universe may still be appreciable in mass and redshift bins where their number density is low.

Such findings mildly vary for different stellar archaeological prescriptions. The most evident difference is in the enhanced contribution from large stellar masses toward high redshift for the \citetalias{Knowles2023} prescription. This can be understood as follows: \citetalias{Knowles2023} assigns very old ages and very peaked star-formation histories to the most massive galaxies (cf. Figure \ref{fig|SAR}), so implying that they are strongly star-forming at high redshift. Although their number density is small (since massive galaxies tend to be rare, and especially so at high $z$; cf. white contours in Figure \ref{fig|SFRD_2D}), the very large SFR of these systems partially compensates their low statistics and allows them to contribute the cosmic SFR density toward redshifts higher than for the other stellar archaeology prescriptions.

\begin{figure}
\epsscale{1}\plotone{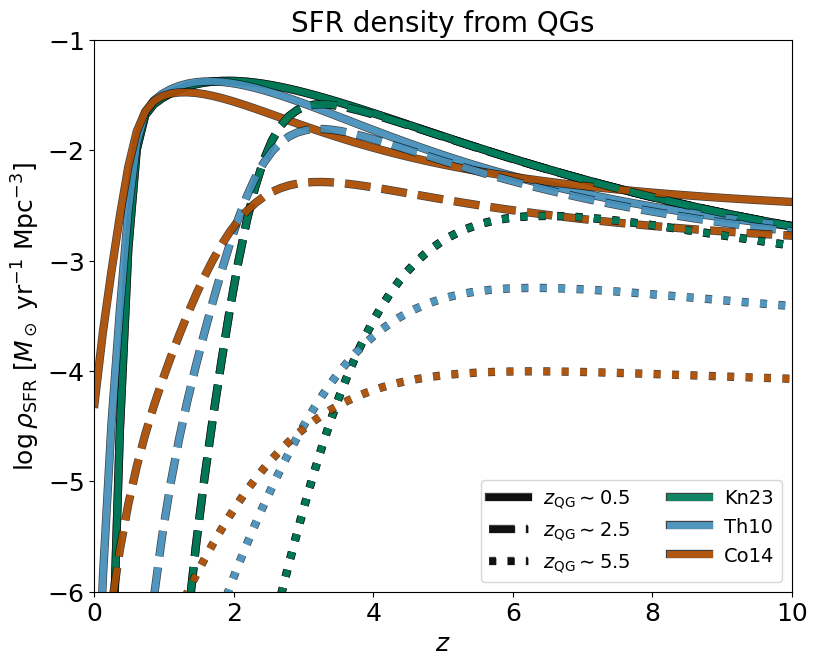}
\caption{The contribution to the cosmic SFR density by progenitors of galaxies that are quiescent at different $z_{\rm QG}\approx 0.5$ (solid), $2.5$ (dashed), $5.5$ (dotted). The green, blue and red lines illustrate the outcomes of \texttt{StAGE} when the stellar archaeology prescriptions by \citetalias{Knowles2023}, \citetalias{Thomas2010} or \citetalias{Conroy2014} are employed, respectively.}\label{fig|SFRD_passz}
\end{figure}

Figure \ref{fig|SFRD_passz} shows the contributions to the SFR density from progenitors of galaxies that are quiescent at different redshifts $z_{\rm QG}$. Since galaxies that are quiescent at higher redshift tend to remain quiescent also at a lower one, the contribution of galaxies quenching at higher $z_{\rm QG}$ is a subset of that from galaxies quenching at lower $z_{\rm QG}$; in other words, the curve with $z_{\rm QG}\approx 0$ includes all the others. Therefore for increasing $z_{\rm QG}$ the SFR density features a peak shifted toward higher redshift and a progressively lower normalization\footnote{Note that for a given $z_{\rm QG}$ the SFR density of galaxies with $z<z_{\rm QG}$ is not exactly null; this is due to the spread in the formation time distribution and galaxy ages, that for consistency with stellar archaeological studies is modeled as a symmetric Gaussian (see Section \ref{sec|methods}).} The dependence of these results on stellar archaeological prescriptions is weak for $z\lesssim 3$, while starts to be evident at higher redshift; this is essentially due to the difference in galaxy age at high relic stellar masses. For example, the ages of \citetalias{Knowles2023} are somewhat higher and this allows a larger number of massive galaxies to get quiescent at high redshift, so enhancing the normalization of the SFR density for increasing $z_{\rm QG}$ with respect to the other prescriptions. 

\subsection{The average assembly history of local QGs}

\begin{figure}
\epsscale{1}\plotone{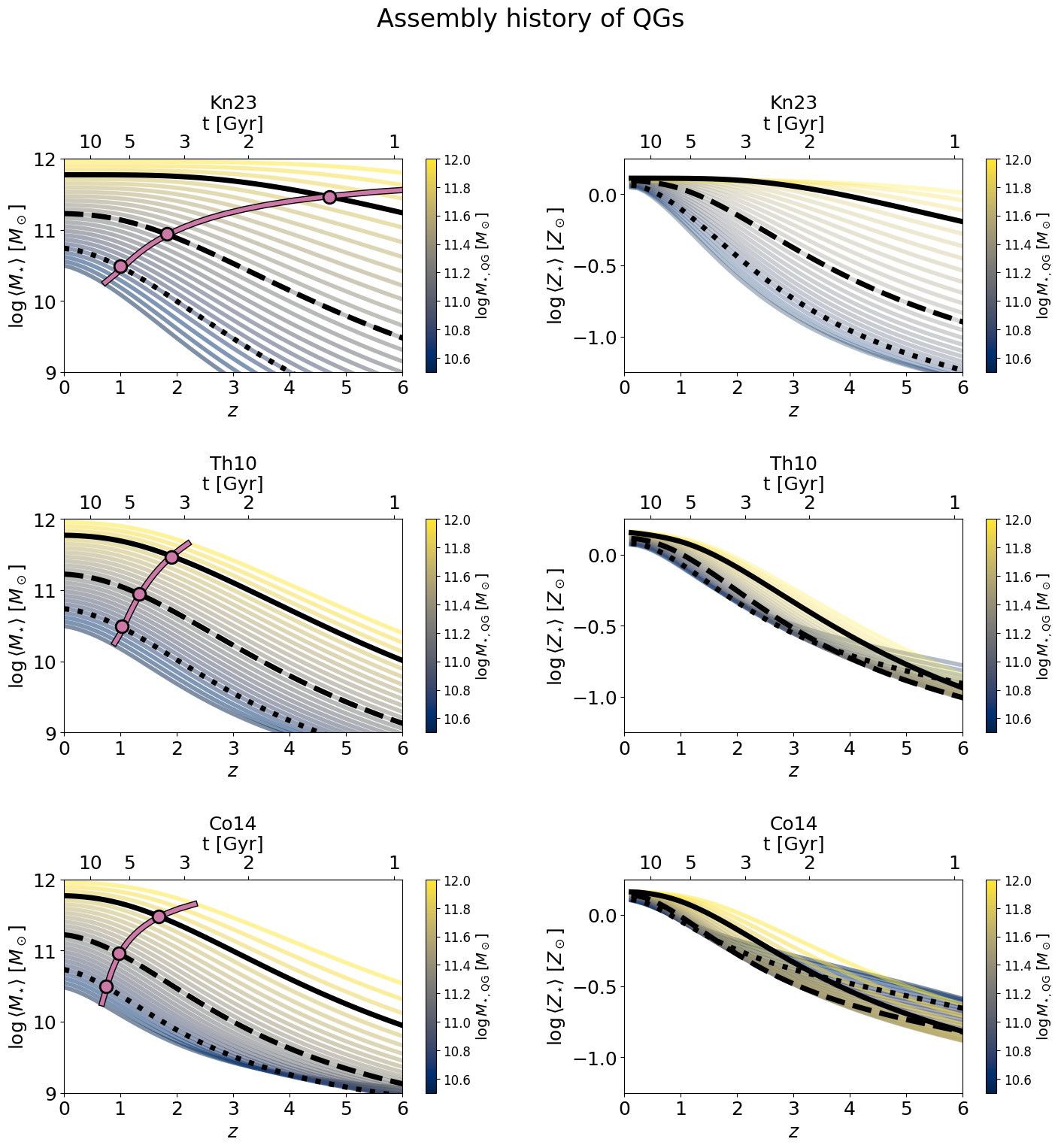}
\caption{The average assembly history of local quiescent galaxies as implied by \texttt{StAGE}. The evolution with redshift (bottom horizontal axis) or cosmic time (top horizontal axis) of the stellar mass (left panels) and of the stellar metallicity (right panels) for the progenitors of local QGs (averaged over the formation time distribution) are illustrated in color scale as a function of the relic stellar mass $M_{\star,\rm QG}$. Top panels refer to the stellar archaeology prescription by \citetalias{Knowles2023}, middle panels to that by \citetalias{Thomas2010} and bottom panels to that by \citetalias{Conroy2014}. 
In all panels the evolutionary tracks for relic stellar masses $\log M_\star\, [M_\odot]\approx 10.75$ (dotted), $11.25$ (dashed) and $11.75$ (solid) are highlighted with black lines. In the left panel, the formation times of the different tracks are connected by a pink line, with the pink dots referring to the three tracks in black.}\label{fig|AHQG}
\end{figure}

In Figure \ref{fig|AHQG} we illustrate the average assembly histories of local QGs as predicted from \texttt{StAGE}. These are meant to be representative of a spatial, time and population averages, while individual star formation histories tend to be less smooth and punctuated by several short-time and spatially-dependent features. We display the evolution with redshift of the stellar mass and stellar metallicity (averaged over the formation time distribution) for galaxies with different relic stellar masses $M_{\star,\rm QG}$ in the range from a few $10^{10}$ to several $10^{11}\, M_\odot$ (color-coded). In the left panels, the pink line illustrates the average formation times of the different mass tracks: this clearly shows the well-known downsizing phenomenon, as the more massive QGs have accumulated most of their stars at earlier epochs. The effect is particularly evident for the \citetalias{Knowles2023} stellar archaeological prescriptions (first row) since the star formation timescale derived from $\alpha-$enhancement is shorter and the age larger, but it is appreciable in all cases, especially when looking at the track evolution in terms of cosmic time (reported in the upper horizontal axis). As to the stellar metallicity, it increases from moderately subsolar values at high redshift to (super)solar ones in the local Universe. Note that when the prescriptions by \citetalias{Thomas2010} or \citetalias{Conroy2014} are employed, the very slow increase of the metal content for low relic masses makes the related evolutionary tracks to cross the ones for more massive QG progenitors at high redshift; this effect is instead not present for the \citetalias{Knowles2023} prescriptions according to which massive systems have a very old age and very peaked star formation histories, to imply that their metal enrichment promptly occurred at very high redshift and then basically stayed put.

\begin{figure}
\epsscale{1}\plotone{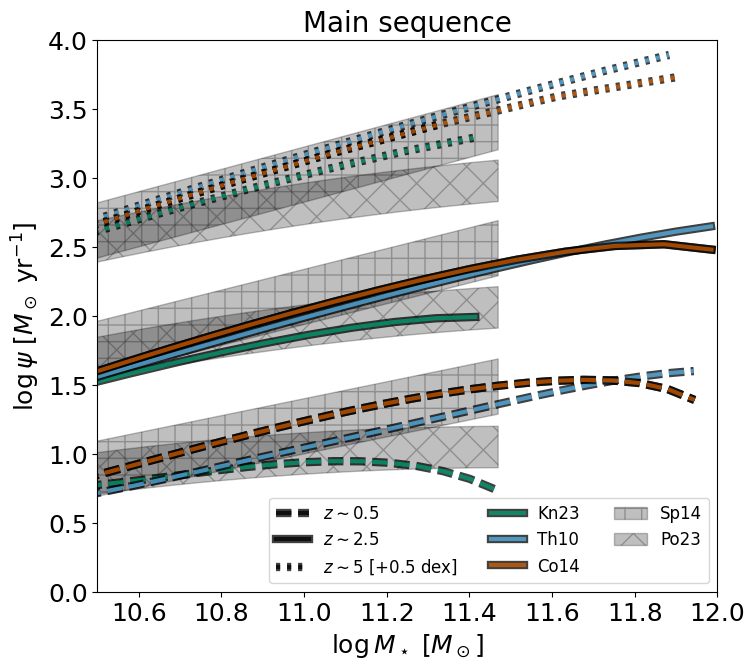}
\caption{The main sequence of star-forming galaxies. Data are from \citet[gray shaded area hatched in crosses]{Popesso2023} and \citet[gray shaded area hatched in plus signs]{Speagle2014}. The green, blue and red lines illustrate the outcomes of \texttt{StAGE} when the stellar archaeology prescriptions by \citetalias{Knowles2023}, \citetalias{Thomas2010} or \citetalias{Conroy2014} are employed, respectively. Dashed lines refer to redshift $z\approx 0.5$, solid to $z\approx 2.5$ and dotted to $z\approx 5$; the results at $z\sim 5$ have been shifted upwards by 0.5 dex for clarity.}\label{fig|MS}
\end{figure}

One could wonder wether during their main star-formation episode the progenitors of local QGs had constituted a peculiar population of objects or not. We clarify this point in Figure \ref{fig|MS}, that shows the galaxy main sequence relationship (SFR vs. stellar mass) for QG progenitors. Dashed, solid and dotted lines illustrate the outcomes of \texttt{StAGE} at different redshifts $z\sim 0.5$, $2.5$ and $5$, respectively, compared with the observational determinations by \cite{Speagle2014} and \cite{Popesso2023}.
All in all, we find that the star-forming progenitors of local QGs are expected to lay on the main sequence. This strongly suggest that most of their stellar mass has been accumulated by in-situ processes. The result is nearly independent on the adopted stellar archaeological prescriptions, as shown by the different colored lines. 

\begin{figure}
\epsscale{1}\plotone{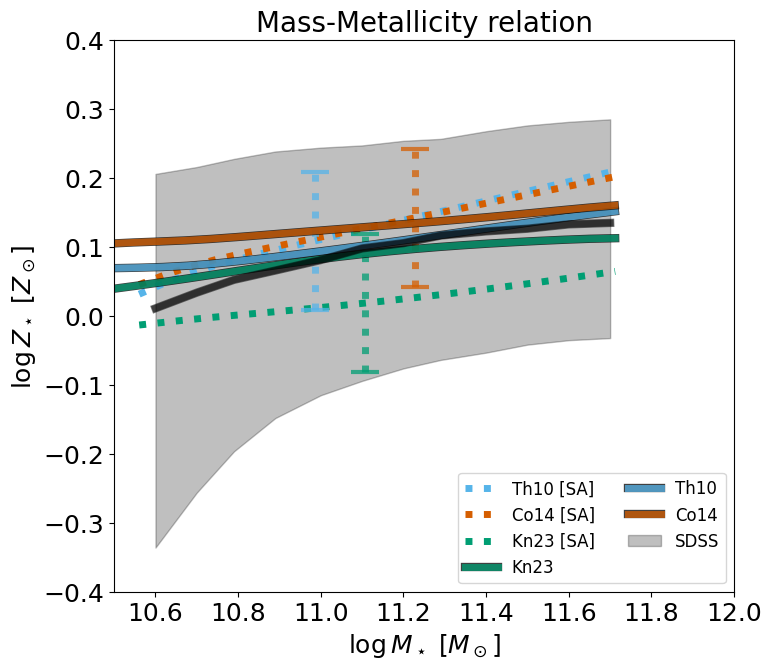}
\caption{The mass-metallicity relationship. Data are from \citet[black solid line illustrates the median, while gray shaded area is the $1\sigma$ uncertainty]{Gallazzi2006}. The green, blue and red lines illustrate the outcomes of \texttt{StAGE} when the stellar archaeology prescriptions by \citetalias{Knowles2023}, \citetalias{Thomas2010} or \citetalias{Conroy2014} are employed, respectively. In particular, solid lines refer to the reconstructed stellar metallicity from our approach that exploits age and $[\alpha/{\rm Fe}]$ from stellar archaeology combined with the FMR. The dotted lines are the relations inferred directly from stellar archaeology, with the dotted errorbars illustrating the related $1\sigma$ uncertainty.}\label{fig|MassMetal}
\end{figure}

In Figure \ref{fig|MassMetal} we show the mass-metallicity relation of local QGs. The average stellar metallicity inferred by \texttt{StAGE}, shown by the colored solid lines, agrees well with the observational determinations by \cite{Gallazzi2006}. More interestingly, we also compare it to the stellar metallicity measured directly from stellar archaeology (dotted lines), which has a typical $1\sigma$ uncertainty around $0.1$ dex (dotted errorbars). This is not a trivial result, since \texttt{StAGE} exploits just the age and [$\alpha$/Fe] vs. stellar mass relation to reconstruct the star-formation histories of local QG; then it computes the gas metallicity in the progenitors of local QGs by the FMR, and eventually averages the latter over the star formation history to get the final stellar metallicity (cf. Equation \ref{eq|metallicity}). On the other hand, in the stellar archaeological approach the stellar metallicity is determined jointly with (it is actually degenerate with) the age and the [$\alpha$/Fe] ratio, via Lick indices or full-spectrum fitting. All in all, it is a remarkable consistency check that the stellar metallicity recovered by \texttt{StAGE} agrees with that from stellar archaeology for the various prescriptions within the related $1\sigma$ uncertainty.

\subsection{How many QGs should we expect at high$-z$?}

Recently, there has been a huge interest in the discovery of a population of relatively massive QGs at $z\sim 2-5$ via JWST observations (e.g., \citealt{Carnall2023,Baker2024}). This seems to challenge cosmological simulations and semi-analytic frameworks (e.g., \citealt{Lagos2024}), although cosmic variance effects could partly account for the discrepancy between data and models. 

\begin{figure}
\epsscale{1}\plotone{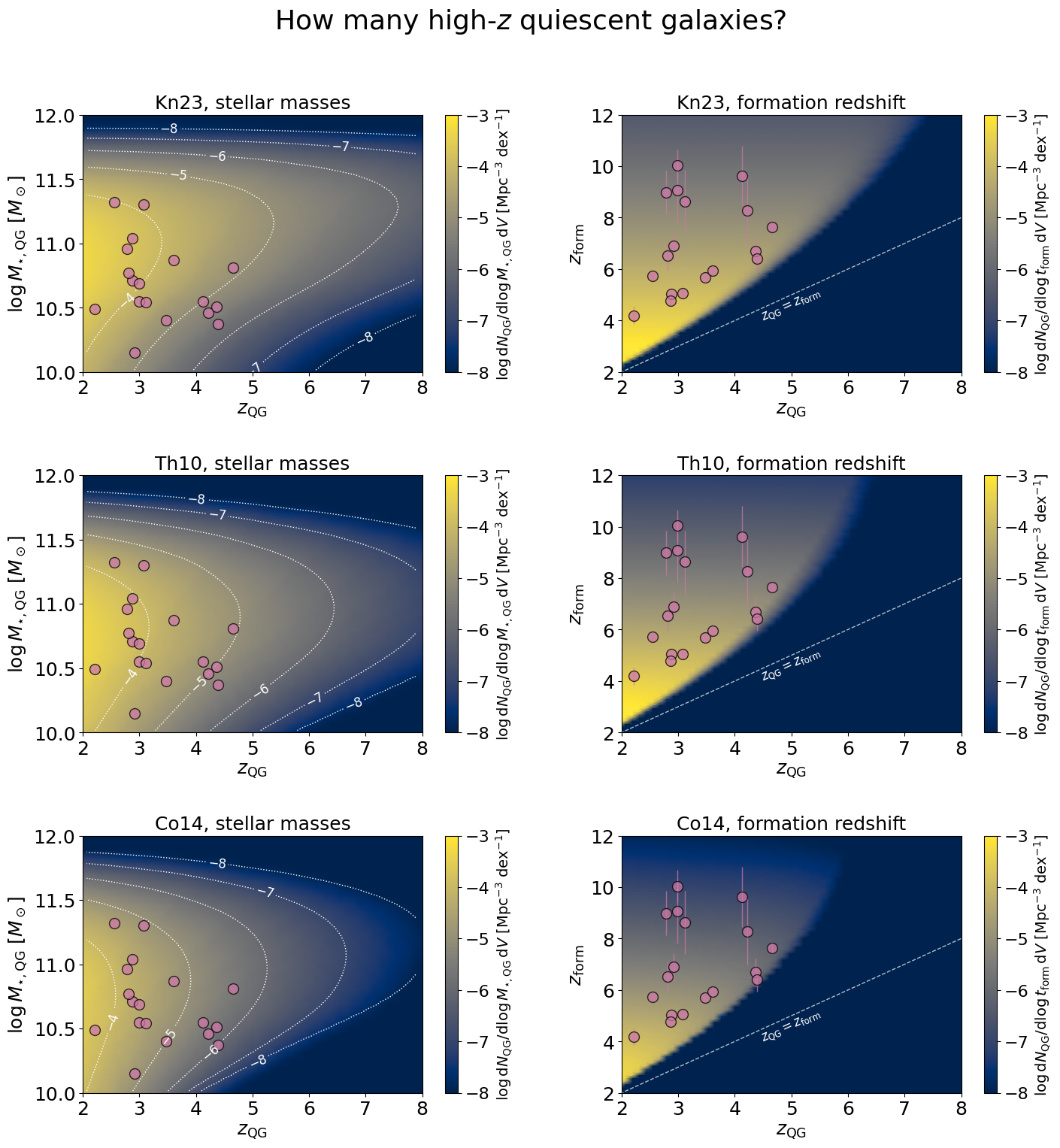}
\caption{The number density of QGs at high redshifts $z_{\rm QG}\gtrsim 2$ sliced in relic stellar mass $M_{\star, \rm QG}$ (left panels) and in formation redshift $z_{\rm form}$ as predicted by \texttt{StAGE} (color-scale and white contours in the left panels). In the right panels the dashed white line marks the locus $z_{\rm QG}=z_{\rm form}$. Top panels refer to the stellar archaeology prescription by \citetalias{Knowles2023}, middle panels to that by \citetalias{Thomas2010} and bottom panels to that by \citetalias{Conroy2014}. Data by \citet[pink circles]{Baker2024} refer to a small (not complete) sample of high-$z$ QG with robust measurements of observed redshifts from spectroscopy and formation redshift estimates from SED fitting}.\label{fig|HighzQG}
\end{figure}

Thus it could be relevant to check the expectation of a framework like \texttt{StAGE}, which is based on completely different data from stellar archaeology, to check whether the observational findings from JWST may suffer of some bias or whether models should instead require substantial revisions.
Figure \ref{fig|HighzQG} displays the number densities of QGs (color-code and white contours) at different $z_{\rm QG}$ and for different relic stellar masses $M_{\star,\rm QG}$ (left panels) or different formation redshifts $z_{\rm form}$ (right panels) as predicted by \texttt{StAGE}. 

Remarkably, \texttt{StAGE} indicates that an appreciable number density of QGs with relic stellar masses $M_{\star,\rm QG}\lesssim$ a few $10^{11}\, M_\odot$ and formation redshifts $z_{\rm form}\lesssim 8-10$ are to be found at $z\sim 2-5$. The density of such quiescent systems is expected to progressively decline only for $z\gtrsim 5$, although one should not be surprised to find some exceptional QGs with relic stellar masses around $10^{11}\, M_\odot$ and formation redshift $z_{\rm form}\gtrsim 12$ at $z_{\rm QG}\sim 8-10$. 

The locations in these diagrams for a sample of high$-z$ QGs identified by JWST (\citealt{Baker2024}), with precise redshift determinations from spectroscopy and estimate of stellar masses and ages (hence formation redshifts) from SED fitting, are signposted by filled pink circles. These datapoints are found in regions of the diagrams where \texttt{StAGE} allows a considerable number of galaxies to be present. In this sense, there is a broad consistency between the JWST data and our model. However, a detailed comparison in terms of number densities is hindered for now since the data by JWST are far from constituting a mass-complete sample in given redshift bins. Future, more extended samples of high-$z$ QGs will enable to further test the predictions of \texttt{StAGE}.

\subsection{Supermassive BH growth and relic masses}

\begin{figure}
\epsscale{1.}\plotone{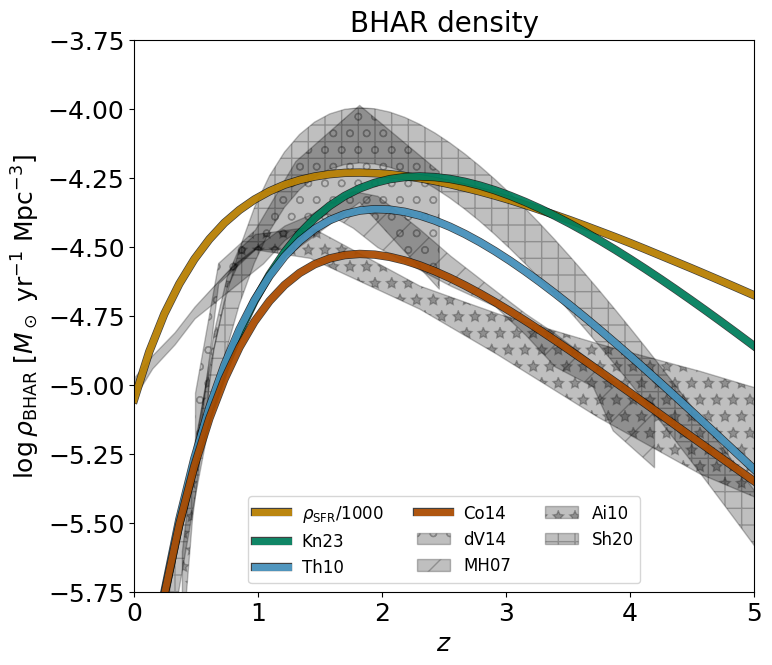}
\caption{The cosmic accretion rate density of supermassive BHs as a function of redshift. The green, blue and red lines illustrate the outcomes of \texttt{StAGE} when the stellar archaeology prescriptions by \citetalias{Knowles2023}, \citetalias{Thomas2010} or \citetalias{Conroy2014} are employed, respectively. The orange line is the cosmic SFR density scaled down by a factor $1000$. Shaded areas display observational determinations by \citet[hatched in circles]{Delvecchio2014}, \citet[hatched in crosses]{Merloni2007}, \citet[hatched in stars]{Aird2010}, and \citet[hatched in plus signs]{Shen2020}.}\label{fig|BHARD}
\end{figure}

Figure \ref{fig|BHARD} illustrates the BH accretion rate density as a function of redshift predicted by \texttt{StAGE} (see Section \ref{sec|BHARD}) according to the different stellar archeological prescriptions (colored lines), compared with various observational estimates by \cite{Merloni2007}, \cite{Aird2010}, \cite{Delvecchio2014} and \cite{Shen2020}. We find a broad agreement with the data, in that the BH accretion rate density peaks at around $z\sim 2-3$, with a tail toward high-redshift that is somewhat dependent on the stellar archaeology prescriptions. At $z\lesssim 0.5$ the BH accretion rate density decreases rapidly because most massive galaxies are already quiescent, and their hosted BH are dead or, in case of reactivations, accrete at very low Eddington ratio. In fact, these reactivations are not included in our modeling, and being associated with jetted emission are mainly pinpointed via radio data.

\begin{figure}
\epsscale{1.}\plotone{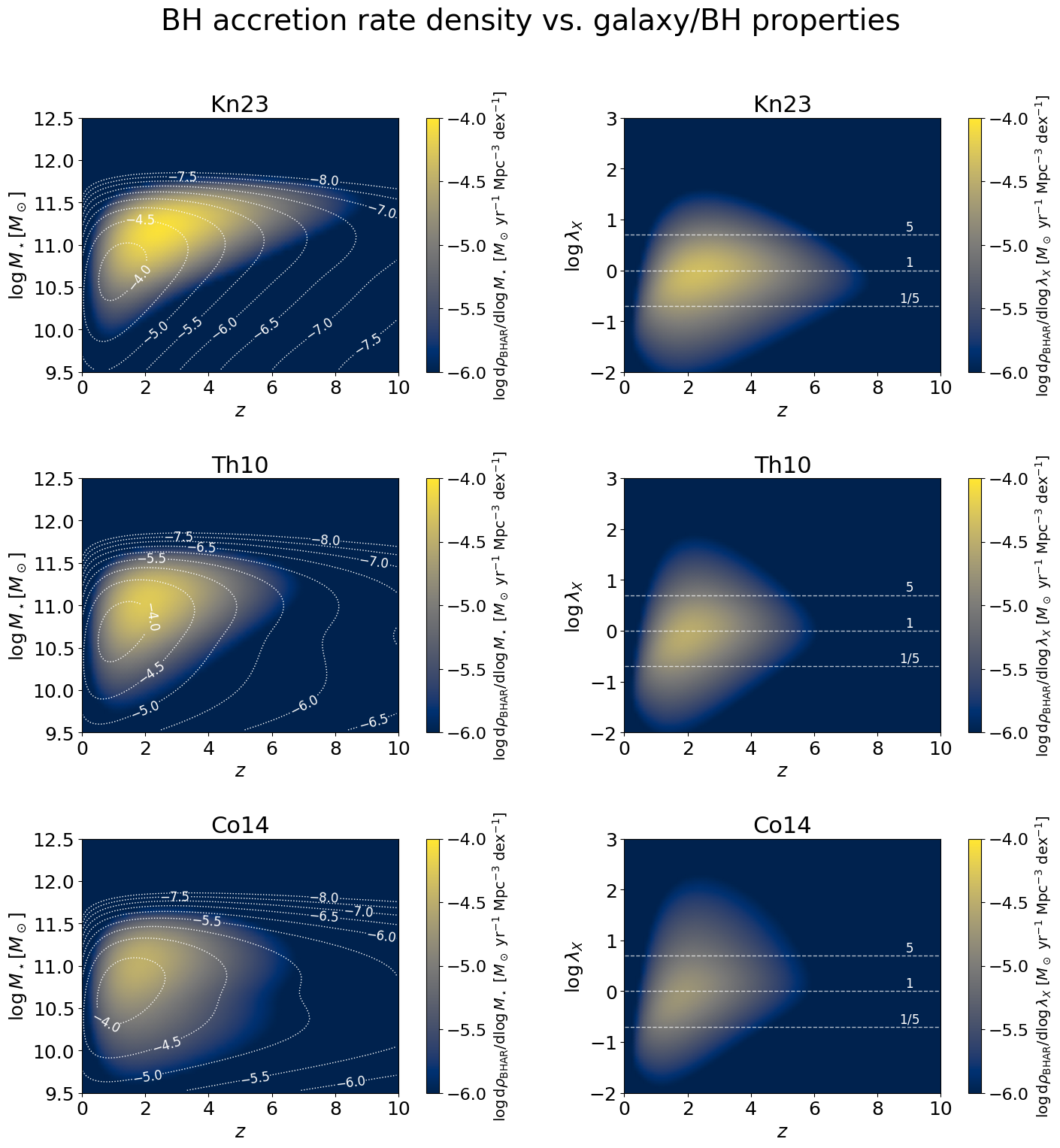}
\caption{The cosmic BH accretion rate density sliced in galaxy stellar mass (left panels) or pseudo-Eddington ratio (right panels) as a function of redshift. Top panels refer to the stellar archaeology prescription by \citetalias{Knowles2023}, middle panels to that by \citetalias{Thomas2010} and bottom panels to that by \citetalias{Conroy2014}. In the left panels the white contours with labels display number densities of galaxies that host accreting supermassive BHs per stellar mass bins $\log [{\rm d}N/{\rm d}\log M_\star\, {\rm d}V]$, in logarithmic units of Mpc$^{-3}$ dex$^{-1}$. In the right panels the three horizontal dashed lines highlight pseudo-Eddington ratio values $\lambda_X = 5$, $1$ and $1/5$ from top to bottom.}\label{fig|BHARD_2d}
\end{figure}

Figure \ref{fig|BHARD_2d} slices the previous result in stellar mass and pseudo-Eddington ratio. It is seen that at high redshift most massive galaxies with $M_\star\gtrsim 10^{11}\, M_\odot$ are the main contributors to the BH accretion rate density, while at lower redshifts $z\lesssim 2-3$ even lower mass galaxies can host accreting BHs. Most of the accretion occurs at pseudo-Eddington ratios close to unity in the range $\lambda_X\sim 1/5-5$, with a tail toward lower values at smaller $z$. These trends are reinforced for the stellar archaeological prescription by \citetalias{Knowles2023} that yields more active BHs in more massive galaxies at higher redshifts. The white contours in the left panels represent the number density of galaxies hosting accreting supermassive BHs. The different shape of
the color pattern with respect to the contours reflects that the former includes a weighting by the accretion rate with its mass, psudo-Eddington ratio and redshift dependence. More massive galaxies are typically hosting more strongly accreting BHs, and this partly offsets the decline in number density to produce a substantial contribution to the cosmic accretion rate density in bins where the number density is low but the stellar mass is high.

\begin{figure}
\epsscale{1}\plotone{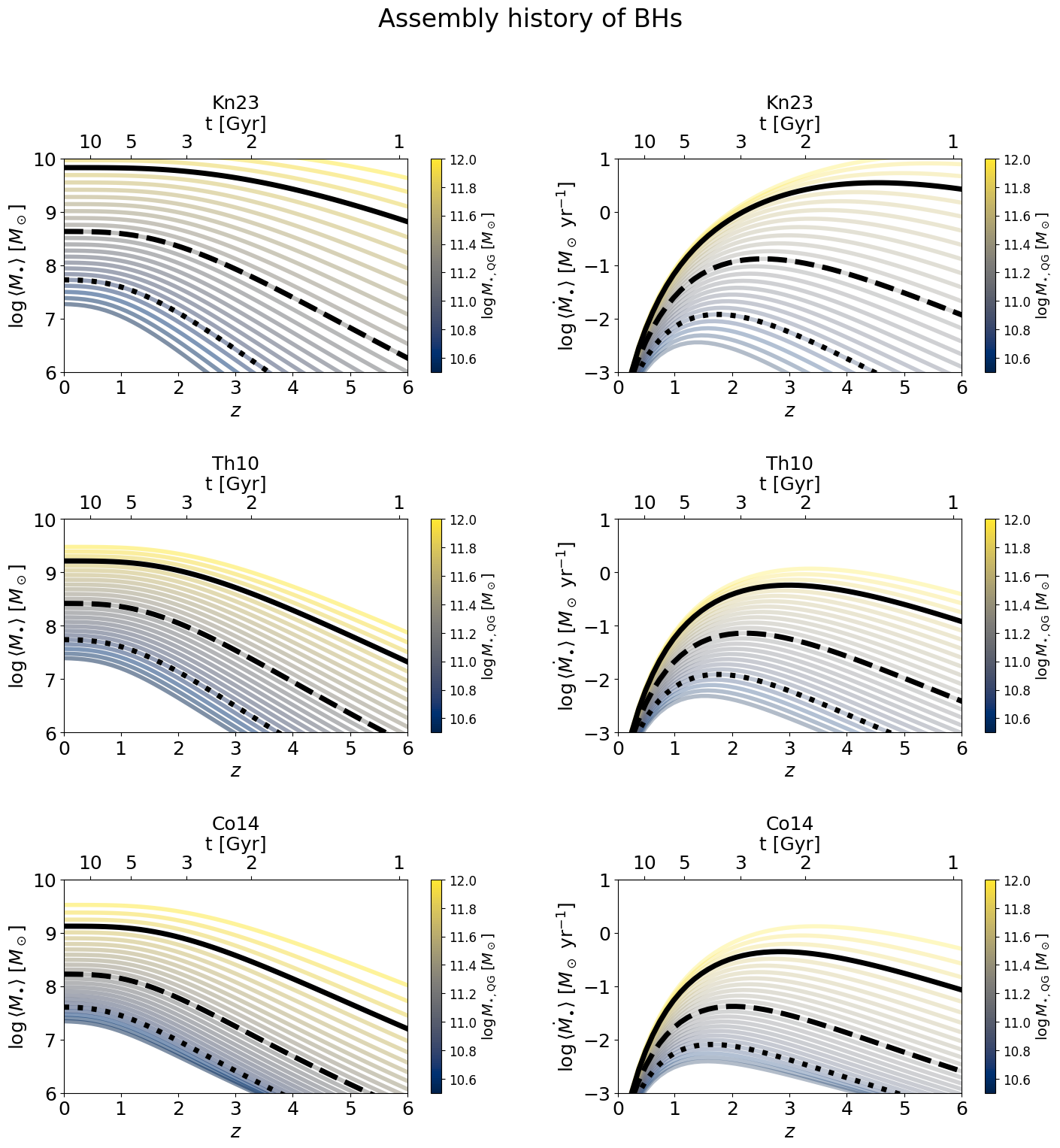}
\caption{The average assembly history of supermassive BHs in the progenitors of local QGs as implied by \texttt{StAGE}. The evolution with redshift (bottom horizontal axis) or cosmic time (top horizontal axis) of the BH mass (left panels) and of the BH accretion rate (right panels) are illustrated in color scale as a function of the relic stellar mass $M_{\star,\rm QG}$. Top panels refer to the stellar archaeology prescription by \citetalias{Knowles2023}, middle panels to that by \citetalias{Thomas2010} and bottom panels to that by \citetalias{Conroy2014}. In all panels the evolutionary tracks for relic stellar masses $\log M_\star\, [M_\odot]\approx 10.75$ (dotted), $11.25$ (dashed) and $11.75$ (solid) are highlighted with black lines.}\label{fig|AHBH}
\end{figure}

Figure \ref{fig|AHBH} illustrates the assembly history of supermassive BHs in the progenitors of local QGs as predicted by \texttt{StAGE}. The left panels display the BH mass, and the right panels the BH accretion rate as a function of redshift (lower horizontal axis) or cosmic time (upper horizontal axis) for different relic stellar masses (color-coded). The accretion rate increases from high redshift and features a broad peak around $z\sim 2-4$ before declining considerably at lower $z$. Correspondingly, the BH mass increases quite rapidly at high redshift and saturates toward the present. The saturation occurs earlier on for more massive galaxies, so that the downsizing phenomenon is evident not only in stellar mass (cf. Figure \ref{fig|AHQG}) but also in the BH component. The detailed growth of the BH mass depends somewhat on the stellar archaeology prescription (different rows in the Figure). In particular, the \citetalias{Knowles2023} prescription tends to yield a very precocious growth of supermassive BHs in galaxies with high relic stellar masses. As already mentioned, this is because \citetalias{Knowles2023} implies early and peaked star-formation histories for massive galaxies. Given that higher BH accretion rates are expected in more star-forming galaxies, the related increase in BH mass is boosted in massive galaxies at early times.

\begin{figure}
\epsscale{1.}\plotone{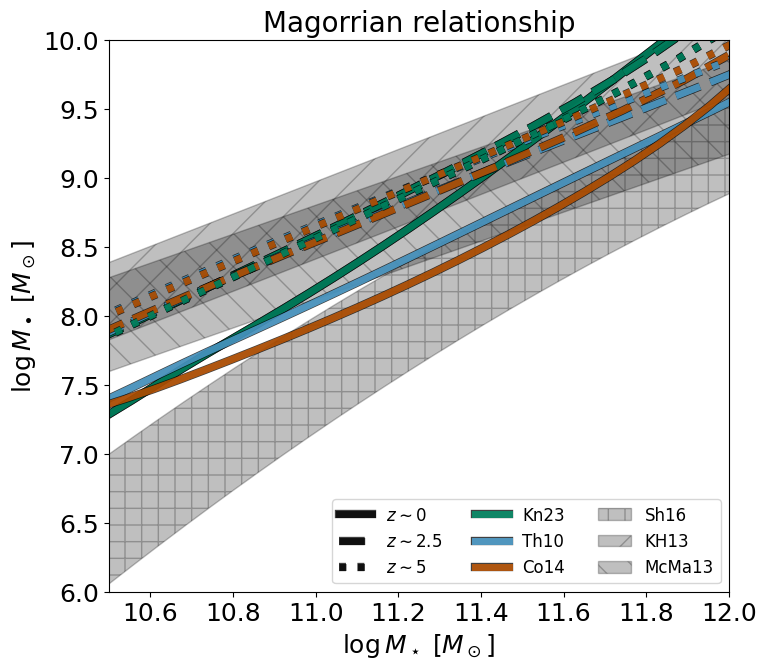}
\caption{The Magorrian relationship between BH and stellar mass. Data at $z\approx 0$ are from \citet[gray shaded area hatched in slashes]{Kormendy2013}, \citet[gray shaded area hatched in backslashes]{McConnell2013}, and \citet[gray shaded area hatched in plus signs]{Shankar2016}. The green, blue and red lines illustrate the outcomes of \texttt{StAGE} when the stellar archaeology prescriptions by \citetalias{Knowles2023}, \citetalias{Thomas2010} or \citetalias{Conroy2014} are employed, respectively. Solid lines refer to redshift $z\approx 0.5$, dashed to $z\approx 2.5$ and dotted to $z\approx 5$.}\label{fig|Magorrian}
\end{figure}

Figure \ref{fig|Magorrian} displays the BH vs. relic stellar mass relation as predicted by \texttt{StAGE} at different redshifts $z\approx 0$ (solid), $2.5$ (dashed) and $5$ (dotted). The local relation is compared with the literature estimates (hatched areas) by \cite{Kormendy2013}, \cite{McConnell2013} and \cite{Shankar2016}, finding a broad agreement within the uncertainties. The outcome is mildly dependent on the adopted stellar archaeological prescription. All in all, the agreement between the Magorrian relation predicted by \texttt{StAGE} with the observed one reinforces the point for a coevolution between the progenitors of QGs and their hosted supermassive BHs.

\section{Summary} \label{sec|summary}

We have built \texttt{StAGE}, a semi-empirical framework of galaxy evolution firmly grounded on stellar archaeology, which is aimed to provide a robust handle on the age, star formation history and population statistics of local QGs and of their progenitors across cosmic times. We have showcased the value of \texttt{StAGE} by addressing some open issues in galaxy and BH evolution. 

\noindent $\bullet$ We have exploited \texttt{StAGE} to compute the cosmic star formation rate (SFR) and stellar mass density contributed by the progenitors of local QGs as a function of redshift, stellar mass and metallicity. We have shown the former quantity to remarkably agree with that estimated for high$-z$ dusty star-forming galaxies which are faint/dark in the NIR, so pointing toward a direct progenitor-descendant
connection among this high-redshift star-forming population and local QGs. Furthermore, we have argued that by appropriately correcting the observed stellar mass density by the contribution of such NIR-dark progenitors, \texttt{StAGE} is able to recover a SFR density which is consistent with direct determinations from UV/IR/radio surveys, so substantially alleviating a longstanding tension between the SFR and stellar mass density estimates. 

\noindent $\bullet$ We have exploited \texttt{StAGE} to compute the metallicity evolution of QGs. Our results support a rapid metal enrichment process in their star-forming progenitors. On average, the abundance may exceed $Z\gtrsim Z_\odot/3$ at the cosmic noon and may already fall in the range $Z\sim Z_\odot/10-Z_\odot/3$ at high redshifts, with only a small fraction of QG progenitors exhibiting $Z\lesssim Z_\odot/10$ up to $z\sim 10$.

\noindent $\bullet$ We have exploited \texttt{StAGE} to provide the average mass assembly history of local QGs, and to show that their star-forming progenitors follow the galaxy main sequence Moreover, stimulated by recent JWST observations of distant QGs, we have also predicted the number of QGs to be expected at high redshift. 

\noindent$\bullet$ We have exploited \texttt{StAGE} to provide the cosmic accretion rate density of supermassive BH as a function of redshift, stellar mass and (pseudo-)Eddington ratio. We have also derived a specific prediction for the Magorrian-like relation between the relic BH and host galaxy stellar masses. 

Many other possible applications of \texttt{StAGE} can be envisaged in the near future. For example, \texttt{StAGE} could be exploited to derive the contribution of QG progenitors to the binary BH cosmic merger rate density as a function of redshift and metallicity of the host galaxy; this may constitute a data-driven way to probabilistically inform codes aimed at searching for the host galaxy of a gravitational wave event, an issue extremely relevant for gravitational wave cosmology, e.g. determination of the Hubble constant from dark sirens. As for cosmological probes, \texttt{StAGE} can also be helpful to refine the `cosmic chronometers' approach, which relies on the aging with redshift of local massive QGs to estimate the Hubble parameter. In the realm of supermassive BHs, the outputs of \texttt{StAGE} can be exploited to derive the population-averaged accretion rate as a function of redshift, BH and galaxy properties; this can be employed in a continuity equation framework to derive the BH mass function at different epochs via a data-driven approach which has not been attempted so far. We also mention the possibility of using \texttt{StAGE} to make detailed predictions concerning the rise of primeval dusty galaxies and quasars toward very high-redshift $z\gtrsim 10$. We plan to pursue some of these possible research developments in forthcoming papers. 

Finally, it would be beneficial to exploit the outputs of \texttt{StAGE} in numerical simulations and semi-analytic frameworks to refine sub-resolution physics, to choose the best parameterizations of the different physical processes at work, and to understand inconsistencies and biases among such ab-initio models and observables. All in all, \texttt{StAGE} may constitute a valuable tool to attack via a data-driven, easily expandable, and computationally low-cost approach the co-evolution of QGs and of their hosted supermassive black holes across cosmic times. 

\begin{acknowledgments}
We thank the anonymous referee for useful comments.
This work was partially funded from the projects: INAF GO-GTO Normal 2023 funding scheme with the project "Serendipitous H-ATLAS-fields Observations of Radio Extragalactic Sources (SHORES)";  INAF Large Grant 2022 funding scheme with the project "MeerKAT and LOFAR Team up: a Unique Radio Window on Galaxy/AGN co-Evolution; ``Data Science methods for MultiMessenger Astrophysics \& Multi-Survey Cosmology'' funded by the Italian Ministry of University and Research, Programmazione triennale 2021/2023 (DM n.2503 dd. 9 December 2019), Programma Congiunto Scuole; Italian Research Center on High Performance Computing Big Data and Quantum Computing (ICSC), project funded by European Union - NextGenerationEU - and National Recovery and Resilience Plan (NRRP) - Mission 4 Component 2 within the activities of Spoke 3 (Astrophysics and Cosmos Observations). MB acknowledges that this article was produced while attending the PhD program in PhD in Space Science and Technology at the University of Trento, Cycle XXXIX, with the support of a scholarship financed by the Ministerial Decree no. 118 of 2nd March 2023, based on the NRRP - funded by the European Union - NextGenerationEU - Mission 4 "Education and Research", Component 1 "Enhancement of the offer of educational services: from nurseries to universities” - Investment 4.1 “Extension of the number of research doctorates and innovative doctorates for public administration and cultural heritage” - CUP E66E23000110001. LB acknowledges financial support from the German Excellence Strategy via the Heidelberg Cluster of Excellence (EXC 2181 - 390900948) STRUCTURES. 
\end{acknowledgments}

\bibliography{ms.bib}

\clearpage

\clearpage

\clearpage

\smallskip

\begin{deluxetable}{lccccccc}
\tablecaption{Marginalized posterior estimates (median and $1\sigma$ confidence intervals, and bestfit in square brackets) of the parameters for a fit to the SFR density vs. redshift.}\label{tab|MCMC}
\tablehead{\colhead{SA Model} & \colhead{$\log N$} & \colhead{$z_{\rm max}$} & \colhead{$\sigma_z$} & \colhead{$\chi_r^2$}}
\startdata
Sh14 & $-1.23_{-0.03}^{+0.03}$ [-1.23] & $1.81_{-0.09}^{+0.09}$ [1.82] & $0.53_{-0.03}^{+0.03}$ [0.53] & $0.61$\\
Sh14 + $\Delta$QG [\citetalias{Knowles2023}] & $-1.14_{-0.03}^{+0.03}$ [-1.14] & $2.29_{-0.09}^{+0.09}$ [2.29] & $0.51_{-0.03}^{+0.03}$ [0.51] & $0.42$\\
Sh14 + $\Delta$QG [\citetalias{Thomas2010}] & $-1.14_{-0.03}^{+0.03}$ [-1.13] & $2.30_{-0.09}^{+0.09}$ [2.30] & $0.51_{-0.03}^{+0.02}$ [0.50] & $0.45$\\
Sh14 + $\Delta$QG [\citetalias{Conroy2014}] & $-1.18_{-0.03}^{+0.03}$ [-1.18] & $2.11_{-0.09}^{+0.09}$ [2.12] & $0.54_{-0.03}^{+0.03}$ [0.53] & $0.48$\\
\enddata
\tablecomments{The SFR density is parametrised assuming a lognormal shape as $\rho_{\rm SFR}(z)=N\,e^{-\ln^2[(1+z)/(1+z_{\rm max})]/2\sigma_z^2}$ , that corresponds to the stellar mass density data of Figure \ref{fig|SMD}. The last column reports the reduced chi-square of the bestfits.}
\end{deluxetable}

\clearpage

\clearpage

\clearpage

\clearpage

\clearpage

\clearpage

\clearpage

\clearpage

\clearpage

\clearpage

\clearpage

\end{document}